\newif\ifDRAFT \DRAFTtrue    
\ifDRAFT
\documentclass[11pt,a4paper,draftcls,onecolumn]{IEEEtran-v17}
\normalsize
\else
\documentclass[10pt, journal]{IEEEtran-v17}
\fi
\pdfoutput=1
\usepackage[english]{babel}
\usepackage{setspace}

\usepackage{xspace}
\usepackage[nosort]{cite}        
\usepackage{url} 
\usepackage[cmex10]{amsmath}
\usepackage{bbm}
\usepackage{graphicx}
\usepackage{paralist}
\usepackage{fancyref}
\usepackage{pdfsync}
\makeatletter
\def\@IEEEinterspaceratioM{0.265}
\def\@IEEEinterspaceMINratioM{0.1651}
\def\@IEEEinterspaceMAXratioM{0.38}

\def\@IEEEinterspaceratioB{0.31}
\def\@IEEEinterspaceMINratioB{0.19}
\def\@IEEEinterspaceMAXratioB{0.38}
\@IEEEtunefonts
\makeatother

\usepackage{amssymb}
\usepackage{amsfonts}
\usepackage{mathrsfs}
\usepackage{xspace}
\usepackage{bm}
\usepackage{upgreek}

\newcommand{\safemath}[2]{\newcommand{#1}{\ensuremath{#2}\xspace}}

\safemath{\bma}{\mathbf{a}}
\safemath{\bmb}{\mathbf{b}}
\safemath{\bmc}{\mathbf{c}}
\safemath{\bmd}{\mathbf{d}}
\safemath{\bme}{\mathbf{e}}
\safemath{\bmf}{\mathbf{f}}
\safemath{\bmg}{\mathbf{g}}
\safemath{\bmh}{\mathbf{h}}
\safemath{\bmi}{\mathbf{i}}
\safemath{\bmj}{\mathbf{j}}
\safemath{\bmk}{\mathbf{k}}
\safemath{\bml}{\mathbf{l}}
\safemath{\bmm}{\mathbf{m}}
\safemath{\bmn}{\mathbf{n}}
\safemath{\bmo}{\mathbf{o}}
\safemath{\bmp}{\mathbf{p}}
\safemath{\bmq}{\mathbf{q}}
\safemath{\bmr}{\mathbf{r}}
\safemath{\bms}{\mathbf{s}}
\safemath{\bmt}{\mathbf{t}}
\safemath{\bmu}{\mathbf{u}}
\safemath{\bmv}{\mathbf{v}}
\safemath{\bmw}{\mathbf{w}}
\safemath{\bmx}{\mathbf{x}}
\safemath{\bmy}{\mathbf{y}}
\safemath{\bmz}{\mathbf{z}}
\safemath{\bmzero}{\mathbf{0}}
\safemath{\bmone}{\mathbf{1}}

\bmdefine{\biad}{a}
\bmdefine{\bibd}{b}
\bmdefine{\bicd}{c}
\bmdefine{\bidd}{d}
\bmdefine{\bied}{e}
\bmdefine{\bifd}{f}
\bmdefine{\bigd}{g}
\bmdefine{\bihd}{h}
\bmdefine{\biid}{i}
\bmdefine{\bijd}{j}
\bmdefine{\bikd}{k}
\bmdefine{\bild}{l}
\bmdefine{\bimd}{m}
\bmdefine{\bind}{n}
\bmdefine{\biod}{o}
\bmdefine{\bipd}{p}
\bmdefine{\biqd}{q}
\bmdefine{\bird}{r}
\bmdefine{\bisd}{s}
\bmdefine{\bitd}{t}
\bmdefine{\biud}{u}
\bmdefine{\bivd}{v}
\bmdefine{\biwd}{w}
\bmdefine{\bixd}{x}
\bmdefine{\biyd}{y}
\bmdefine{\bizd}{z}

\bmdefine{\bixid}{\xi}
\bmdefine{\bilambdad}{\lambda}
\bmdefine{\bimud}{\mu}
\bmdefine{\bithetad}{\theta}
\bmdefine{\biphid}{\phi}

\safemath{\bmia}{\biad}
\safemath{\bmib}{\bibd}
\safemath{\bmic}{\bicd}
\safemath{\bmid}{\bidd}
\safemath{\bmie}{\bied}
\safemath{\bmif}{\bifd}
\safemath{\bmig}{\bigd}
\safemath{\bmih}{\bihd}
\safemath{\bmii}{\biid}
\safemath{\bmij}{\bijd}
\safemath{\bmik}{\bikd}
\safemath{\bmil}{\bild}
\safemath{\bmim}{\bimd}
\safemath{\bmin}{\bind}
\safemath{\bmio}{\biod}
\safemath{\bmip}{\bipd}
\safemath{\bmiq}{\biqd}
\safemath{\bmir}{\bird}
\safemath{\bmis}{\bisd}
\safemath{\bmit}{\bitd}
\safemath{\bmiu}{\biud}
\safemath{\bmiv}{\bivd}
\safemath{\bmiw}{\biwd}
\safemath{\bmix}{\bixd}
\safemath{\bmiy}{\biyd}
\safemath{\bmiz}{\bizd}

\safemath{\bmxi}{\bixid}
\safemath{\bmlambda}{\bilambdad}
\safemath{\bmmu}{\bimud}
\safemath{\bmtheta}{\bithetad}
\safemath{\bmphi}{\biphid}

\safemath{\bA}{\mathbf{A}}
\safemath{\bB}{\mathbf{B}}
\safemath{\bC}{\mathbf{C}}
\safemath{\bD}{\mathbf{D}}
\safemath{\bE}{\mathbf{E}}
\safemath{\bF}{\mathbf{F}}
\safemath{\bG}{\mathbf{G}}
\safemath{\bH}{\mathbf{H}}
\safemath{\bI}{\mathbf{I}}
\safemath{\bJ}{\mathbf{J}}
\safemath{\bK}{\mathbf{K}}
\safemath{\bL}{\mathbf{L}}
\safemath{\bM}{\mathbf{M}}
\safemath{\bN}{\mathbf{N}}
\safemath{\bO}{\mathbf{O}}
\safemath{\bP}{\mathbf{P}}
\safemath{\bQ}{\mathbf{Q}}
\safemath{\bR}{\mathbf{R}}
\safemath{\bS}{\mathbf{S}}
\safemath{\bT}{\mathbf{T}}
\safemath{\bU}{\mathbf{U}}
\safemath{\bV}{\mathbf{V}}
\safemath{\bW}{\mathbf{W}}
\safemath{\bX}{\mathbf{X}}
\safemath{\bY}{\mathbf{Y}}
\safemath{\bZ}{\mathbf{Z}}

\safemath{\bZero}{\mathbf{0}}
\safemath{\bOne}{\mathbf{1}}
\safemath{\bDelta}{\mathbf{\Delta}}
\safemath{\bLambda}{\mathbf{\UpLambda}}
\safemath{\bPhi}{\mathbf{\Upphi}}
\safemath{\bSigma}{\mathbf{\Upsigma}}
\safemath{\bOmega}{\mathbf{\Upomega}}
\safemath{\bTheta}{\mathbf{\Uptheta}}

\bmdefine{\biAd}{A}
\bmdefine{\biBd}{B}
\bmdefine{\biCd}{C}
\bmdefine{\biDd}{D}
\bmdefine{\biEd}{E}
\bmdefine{\biFd}{F}
\bmdefine{\biGd}{G}
\bmdefine{\biHd}{H}
\bmdefine{\biId}{I}
\bmdefine{\biJd}{J}
\bmdefine{\biKd}{K}
\bmdefine{\biLd}{L}
\bmdefine{\biMd}{M}
\bmdefine{\biOd}{N}
\bmdefine{\biPd}{O}
\bmdefine{\biQd}{P}
\bmdefine{\biRd}{R}
\bmdefine{\biSd}{S}
\bmdefine{\biTd}{T}
\bmdefine{\biUd}{U}
\bmdefine{\biVd}{V}
\bmdefine{\biWd}{W}
\bmdefine{\biXd}{X}
\bmdefine{\biYd}{Y}
\bmdefine{\biZd}{Z}

\bmdefine{\biDelta}{\Delta}
\bmdefine{\biLambda}{\Lambda}
\bmdefine{\biPhi}{\Phi}
\bmdefine{\biSigma}{\Sigma}
\bmdefine{\biOmega}{\Omega}
\bmdefine{\biTheta}{\Theta}

\safemath{\bimA}{\biAd}
\safemath{\bimB}{\biBd}
\safemath{\bimC}{\biCd}
\safemath{\bimD}{\biDd}
\safemath{\bimE}{\biEd}
\safemath{\bimF}{\biFd}
\safemath{\bimG}{\biGd}
\safemath{\bimH}{\biHd}
\safemath{\bimI}{\biId}
\safemath{\bimJ}{\biJd}
\safemath{\bimK}{\biKd}
\safemath{\bimL}{\biLd}
\safemath{\bimM}{\biMd}
\safemath{\bimN}{\biNd}
\safemath{\bimO}{\biOd}
\safemath{\bimP}{\biPd}
\safemath{\bimQ}{\biQd}
\safemath{\bimR}{\biRd}
\safemath{\bimS}{\biSd}
\safemath{\bimT}{\biTd}
\safemath{\bimU}{\biUd}
\safemath{\bimV}{\biVd}
\safemath{\bimW}{\biWd}
\safemath{\bimX}{\biXd}
\safemath{\bimY}{\biYd}
\safemath{\bimZ}{\biZd}

\safemath{\bimDelta}{\biDelta}
\safemath{\bimLambda}{\biLambda}
\safemath{\bimPhi}{\biPhi}
\safemath{\bimSigma}{\biSigma}
\safemath{\bimOmega}{\biOmega}
\safemath{\bimTheta}{\biTheta}

\safemath{\setA}{\mathcal{A}}
\safemath{\setB}{\mathcal{B}}
\safemath{\setC}{\mathcal{C}}
\safemath{\setD}{\mathcal{D}}
\safemath{\setE}{\mathcal{E}}
\safemath{\setF}{\mathcal{F}}
\safemath{\setG}{\mathcal{G}}
\safemath{\setH}{\mathcal{H}}
\safemath{\setI}{\mathcal{I}}
\safemath{\setJ}{\mathcal{J}}
\safemath{\setK}{\mathcal{K}}
\safemath{\setL}{\mathcal{L}}
\safemath{\setM}{\mathcal{M}}
\safemath{\setN}{\mathcal{N}}
\safemath{\setO}{\mathcal{O}}
\safemath{\setP}{\mathcal{P}}
\safemath{\setQ}{\mathcal{Q}}
\safemath{\setR}{\mathcal{R}}
\safemath{\setS}{\mathcal{S}}
\safemath{\setT}{\mathcal{T}}
\safemath{\setU}{\mathcal{U}}
\safemath{\setV}{\mathcal{V}}
\safemath{\setW}{\mathcal{W}}
\safemath{\setX}{\mathcal{X}}
\safemath{\setY}{\mathcal{Y}}
\safemath{\setZ}{\mathcal{Z}}
\safemath{\emptySet}{\varnothing}

\safemath{\colA}{\mathscr{A}}
\safemath{\colB}{\mathscr{B}}
\safemath{\colC}{\mathscr{C}}
\safemath{\colD}{\mathscr{D}}
\safemath{\colE}{\mathscr{E}}
\safemath{\colF}{\mathscr{F}}
\safemath{\colG}{\mathscr{G}}
\safemath{\colH}{\mathscr{H}}
\safemath{\colI}{\mathscr{I}}
\safemath{\colJ}{\mathscr{J}}
\safemath{\colK}{\mathscr{K}}
\safemath{\colL}{\mathscr{L}}
\safemath{\colM}{\mathscr{M}}
\safemath{\colN}{\mathscr{N}}
\safemath{\colO}{\mathscr{O}}
\safemath{\colP}{\mathscr{P}}
\safemath{\colQ}{\mathscr{Q}}
\safemath{\colR}{\mathscr{R}}
\safemath{\colS}{\mathscr{S}}
\safemath{\colT}{\mathscr{T}}
\safemath{\colU}{\mathscr{U}}
\safemath{\colV}{\mathscr{V}}
\safemath{\colW}{\mathscr{W}}
\safemath{\colX}{\mathscr{X}}
\safemath{\colY}{\mathscr{Y}}
\safemath{\colZ}{\mathscr{Z}}

\safemath{\opA}{\mathbb{A}}
\safemath{\opB}{\mathbb{B}}
\safemath{\opC}{\mathbb{C}}
\safemath{\opD}{\mathbb{D}}
\safemath{\opE}{\mathbb{E}}
\safemath{\opF}{\mathbb{F}}
\safemath{\opG}{\mathbb{G}}
\safemath{\opH}{\mathbb{H}}
\safemath{\opI}{\mathbb{I}}
\safemath{\opJ}{\mathbb{J}}
\safemath{\opK}{\mathbb{K}}
\safemath{\opL}{\mathbb{L}}
\safemath{\opM}{\mathbb{M}}
\safemath{\opN}{\mathbb{N}}
\safemath{\opO}{\mathbb{O}}
\safemath{\opP}{\mathbb{P}}
\safemath{\opQ}{\mathbb{Q}}
\safemath{\opR}{\mathbb{R}}
\safemath{\opS}{\mathbb{S}}
\safemath{\opT}{\mathbb{T}}
\safemath{\opU}{\mathbb{U}}
\safemath{\opV}{\mathbb{V}}
\safemath{\opW}{\mathbb{W}}
\safemath{\opX}{\mathbb{X}}
\safemath{\opY}{\mathbb{Y}}
\safemath{\opZ}{\mathbb{Z}}
\safemath{\opZero}{\mathbb{O}}
\safemath{\identityop}{\opI}

\safemath{\veca}{\bma}
\safemath{\vecb}{\bmb}
\safemath{\vecc}{\bmc}
\safemath{\vecd}{\bmd}
\safemath{\vece}{\bme}
\safemath{\vecf}{\bmf}
\safemath{\vecg}{\bmg}
\safemath{\vech}{\bmh}
\safemath{\veci}{\bmi}
\safemath{\vecj}{\bmj}
\safemath{\veck}{\bmk}
\safemath{\vecl}{\bml}
\safemath{\vecm}{\bmm}
\safemath{\vecn}{\bmn}
\safemath{\veco}{\bmo}
\safemath{\vecp}{\bmp}
\safemath{\vecq}{\bmq}
\safemath{\vecr}{\bmr}
\safemath{\vecs}{\bms}
\safemath{\vect}{\bmt}
\safemath{\vecu}{\bmu}
\safemath{\vecv}{\bmv}
\safemath{\vecw}{\bmw}
\safemath{\vecx}{\bmx}
\safemath{\vecy}{\bmy}
\safemath{\vecz}{\bmz}

\safemath{\veczero}{\bmzero}
\safemath{\vecone}{\bmone}
\safemath{\vecxi}{\bmxi}
\safemath{\veclambda}{\bmlambda}
\safemath{\vecmu}{\bmmu}
\safemath{\vectheta}{\bmtheta}
\safemath{\vecphi}{\bmphi}

\safemath{\matA}{\bA}
\safemath{\matB}{\bB}
\safemath{\matC}{\bC}
\safemath{\matD}{\bD}
\safemath{\matE}{\bE}
\safemath{\matF}{\bF}
\safemath{\matG}{\bG}
\safemath{\matH}{\bH}
\safemath{\matI}{\bI}
\safemath{\matJ}{\bJ}
\safemath{\matK}{\bK}
\safemath{\matL}{\bL}
\safemath{\matM}{\bM}
\safemath{\matN}{\bN}
\safemath{\matO}{\bO}
\safemath{\matP}{\bP}
\safemath{\matQ}{\bQ}
\safemath{\matR}{\bR}
\safemath{\matS}{\bS}
\safemath{\matT}{\bT}
\safemath{\matU}{\bU}
\safemath{\matV}{\bV}
\safemath{\matW}{\bW}
\safemath{\matX}{\bX}
\safemath{\matY}{\bY}
\safemath{\matZ}{\bZ}
\safemath{\matzero}{\bmzero}

\safemath{\matDelta}{\bDelta}
\safemath{\matLambda}{\bLambda}
\safemath{\matPhi}{\bPhi}
\safemath{\matSigma}{\bSigma}
\safemath{\matOmega}{\bOmega}
\safemath{\matTheta}{\bTheta}

\safemath{\matidentity}{\matI}
\safemath{\matone}{\matO}

\safemath{\rnda}{A}
\safemath{\rndb}{B}
\safemath{\rndc}{C}
\safemath{\rndd}{D}
\safemath{\rnde}{E}
\safemath{\rndf}{F}
\safemath{\rndg}{G}
\safemath{\rndh}{H}
\safemath{\rndi}{I}
\safemath{\rndj}{J}
\safemath{\rndk}{K}
\safemath{\rndl}{L}
\safemath{\rndm}{M}
\safemath{\rndn}{N}
\safemath{\rndo}{O}
\safemath{\rndp}{P}
\safemath{\rndq}{Q}
\safemath{\rndr}{R}
\safemath{\rnds}{S}
\safemath{\rndt}{T}
\safemath{\rndu}{U}
\safemath{\rndv}{V}
\safemath{\rndw}{W}
\safemath{\rndx}{X}
\safemath{\rndy}{Y}
\safemath{\rndz}{Z}

\safemath{\rveca}{\bimA}
\safemath{\rvecb}{\bimB}
\safemath{\rvecc}{\bimC}
\safemath{\rvecd}{\bimD}
\safemath{\rvece}{\bimE}
\safemath{\rvecf}{\bimF}
\safemath{\rvecg}{\bimG}
\safemath{\rvech}{\bimH}
\safemath{\rveci}{\bimI}
\safemath{\rvecj}{\bimJ}
\safemath{\rveck}{\bimK}
\safemath{\rvecl}{\bimL}
\safemath{\rvecm}{\bimM}
\safemath{\rvecn}{\bimN}
\safemath{\rveco}{\bomO}
\safemath{\rvecp}{\bimP}
\safemath{\rvecq}{\bimQ}
\safemath{\rvecr}{\bimR}
\safemath{\rvecs}{\bimS}
\safemath{\rvect}{\bimT}
\safemath{\rvecu}{\bimU}
\safemath{\rvecv}{\bimV}
\safemath{\rvecw}{\bimW}
\safemath{\rvecx}{\bimX}
\safemath{\rvecy}{\bimY}
\safemath{\rvecz}{\bimZ}

\safemath{\rvecxi}{\bmxi}
\safemath{\rveclambda}{\bmlambda}
\safemath{\rvecmu}{\bmmu}
\safemath{\rvectheta}{\bmtheta}
\safemath{\rvecphi}{\bmphi}

\safemath{\rmatA}{\bimA}
\safemath{\rmatB}{\bimB}
\safemath{\rmatC}{\bimC}
\safemath{\rmatD}{\bimD}
\safemath{\rmatE}{\bimE}
\safemath{\rmatF}{\bimF}
\safemath{\rmatG}{\bimG}
\safemath{\rmatH}{\bimH}
\safemath{\rmatI}{\bimI}
\safemath{\rmatJ}{\bimJ}
\safemath{\rmatK}{\bimK}
\safemath{\rmatL}{\bimL}
\safemath{\rmatM}{\bimM}
\safemath{\rmatN}{\bimN}
\safemath{\rmatO}{\bimO}
\safemath{\rmatP}{\bimP}
\safemath{\rmatQ}{\bimQ}
\safemath{\rmatR}{\bimR}
\safemath{\rmatS}{\bimS}
\safemath{\rmatT}{\bimT}
\safemath{\rmatU}{\bimU}
\safemath{\rmatV}{\bimV}
\safemath{\rmatW}{\bimW}
\safemath{\rmatX}{\bimX}
\safemath{\rmatY}{\bimY}
\safemath{\rmatZ}{\bimZ}

\safemath{\rmatDelta}{\bimDelta}
\safemath{\rmatLambda}{\bimLambda}
\safemath{\rmatPhi}{\bimPhi}
\safemath{\rmatSigma}{\bimSigma}
\safemath{\rmatOmega}{\bimOmega}
\safemath{\rmatTheta}{\bimTheta}

\usepackage{amssymb}
\usepackage{amsfonts}
\usepackage{mathrsfs}
\usepackage{xspace}
\usepackage{bm}
\usepackage{fancyref}
\usepackage{textcomp}

\newenvironment{textbmatrix}{	\setlength{\arraycolsep}{2.5pt}%
								\big[\begin{matrix}}{\end{matrix}\big]%
								\raisebox{0.08ex}{\vphantom{M}}}

\def\be{\begin{equation}}
\def\ee{\end{equation}}
\def\een{\nonumber \end{equation}}
\def\mat{\begin{bmatrix}}
\def\emat{\end{bmatrix}}
\def\btm{\begin{textbmatrix}}
\def\etm{\end{textbmatrix}}

\def\ba#1\ea{\begin{align}#1\end{align}}
\def\bas#1\eas{\begin{align*}#1\end{align*}}
\def\bs#1\es{\begin{split}#1\end{split}} 
\def\bg#1\eg{\begin{gather}#1\end{gather}}
\def\bml#1\eml{\begin{multline}#1\end{multline}}
\def\bi#1\ei{\begin{itemize}#1\end{itemize}}

\newcommand{\lefto}{\mathopen{}\left}

\DeclareMathOperator{\rank}{rank}			
\DeclareMathOperator{\Prob}{\opP}			
\DeclareMathOperator{\Exop}{\opE}			

\newcommand{\abs}[1]{\lefto\lvert#1\right\rvert}		

\newcommand{\vecnorm}[1]{\lefto\lVert#1\right\rVert}		
\newcommand{\tp}[1]{\ensuremath{#1^{T}}} 		
\newcommand{\herm}[1]{\ensuremath{#1^{H}}} 	

\safemath{\dirac}{\delta}					
\safemath{\krond}{\dirac}					
\newcommand{\allo}[2]{\ensuremath{#1=1,2,\ldots,#2}}
\newcommand{\allonotwo}[2]{\ensuremath{#1=1,\ldots,#2}}

\safemath{\upto}{\uparrow}
\safemath{\downto}{\downarrow}
\safemath{\iu}{j}							
\safemath{\ev}{\lambda}						
\safemath{\hilseqspace}{l^{2}}				
\newcommand{\banachfunspace}[1]{\setL^{#1}}	
\safemath{\hilfunspace}{\banachfunspace{2}}	

\safemath{\SNR}{\text{\sc snr}} 				
\safemath{\No}{N_0}							
\safemath{\Es}{E_s}							
\safemath{\Eb}{E_b}							
\safemath{\EbNo}{\frac{\Eb}{\No}}
\safemath{\EsNo}{\frac{\Es}{\No}}

\DeclareMathOperator{\CHop}{\ensuremath{\opH}} 
\safemath{\tvir}{\rndh_{\CHop}}				
\safemath{\tvtf}{\rndl_{\CHop}}				
\safemath{\spf}{\rnds_{\CHop}}				
\safemath{\bff}{H_{\CHop}}					

\safemath{\ircf}{r_{h}}						
\safemath{\tftvcf}{r_{s}}					
\safemath{\tfcf}{r_{l}}						
\safemath{\bfcf}{r_{H}}						

\safemath{\tcorr}{c_h}						
\safemath{\scf}{c_{s}}						
\safemath{\tfcorr}{c_{l}}					
\safemath{\fcorr}{c_{H}}						

\safemath{\mi}{I}							
\safemath{\capacity}{C}						

\safemath{\normal}{\mathcal{N}}			
\safemath{\jpg}{\mathcal{CN}}			
\safemath{\mchain}{\leftrightarrow}		

\safemath{\dB}{\,\mathrm{dB}}
\safemath{\dBm}{\,\mathrm{dBm}}
\safemath{\Hz}{\,\mathrm{Hz}}
\safemath{\kHz}{\,\mathrm{kHz}}
\safemath{\MHz}{\,\mathrm{MHz}}
\safemath{\GHz}{\,\mathrm{GHz}}
\safemath{\s}{\,\mathrm{s}}
\safemath{\ms}{\,\mathrm{ms}}
\safemath{\mus}{\,\mathrm{\text{\textmu}s}}
\safemath{\ns}{\,\mathrm{ns}}
\safemath{\ps}{\,\mathrm{ps}}
\safemath{\meter}{\,\mathrm{m}}
\safemath{\mm}{\,\mathrm{mm}}
\safemath{\cm}{\,\mathrm{cm}}
\safemath{\m}{\,\mathrm{m}}
\safemath{\W}{\,\mathrm{W}}
\safemath{\mW}{\, \mathrm{mW}}
\safemath{\J}{\,\mathrm{J}}
\safemath{\K}{\,\mathrm{K}}
\safemath{\bit}{\,\mathrm{bit}}
\safemath{\nat}{\,\mathrm{nat}}

\safemath{\define}{=}			

\safemath{\equivalent}{\sim}
\safemath{\distas}{\sim}					
\safemath{\sdiff}{\Delta}				

\safemath{\reals}{\mathbb{R}}
\safemath{\positivereals}{\reals_{+}}
\safemath{\integers}{\mathbb{Z}}
\safemath{\posint}{\integers_{+}}
\safemath{\naturals}{\mathbb{N}}
\safemath{\posnaturals}{\naturals_{+}}
\safemath{\complexset}{\mathbb{C}}
\safemath{\rationals}{\mathbb{Q}}

\newcommand*{\fancyrefapplabelprefix}{app}		
\newcommand*{\fancyrefthmlabelprefix}{thm}		
\newcommand*{\fancyreflemlabelprefix}{lem}		
\newcommand*{\fancyrefcorlabelprefix}{cor}		
\newcommand*{\fancyrefdeflabelprefix}{def}		
\newcommand*{\fancyrefproplabelprefix}{prop}		
\frefformat{vario}{\fancyrefseclabelprefix}{Section~#1}
\frefformat{vario}{\fancyrefthmlabelprefix}{Theorem~#1}
\frefformat{vario}{\fancyreflemlabelprefix}{Lemma~#1}
\frefformat{vario}{\fancyrefcorlabelprefix}{Corollary~#1}
\frefformat{vario}{\fancyrefdeflabelprefix}{Definition~#1}
\frefformat{vario}{\fancyreffiglabelprefix}{Fig.~#1}
\frefformat{vario}{\fancyrefapplabelprefix}{Appendix~#1}
\frefformat{vario}{\fancyrefeqlabelprefix}{(#1)}
\frefformat{vario}{\fancyrefproplabelprefix}{Property~#1}

 \newcommand{\trm}{\textrm}
 \newcommand{\dummyrel}[1]{\mathrel{\hphantom{#1}}\strut\mskip-\medmuskip}
 \newtheorem{thm}{Theorem}
 \newtheorem{cor}[thm]{Corollary}

 \newtheorem{lem}[thm]{Lemma}

\safemath{\dict}{\matD}
\safemath{\inputdim}{N}		
\safemath{\outputdim}{M}		
\safemath{\sparsity}{S}	
\safemath{\inputdimA}{{N_a}}	
\safemath{\inputdimB}{{N_b}}	
\safemath{\elemA}{{n_a}}	
\safemath{\elemB}{{n_b}}	
\safemath{\resA}{\matR_a}	
\safemath{\resB}{\matR_b}	
\safemath{\subD}{\matS} 
\safemath{\subA}{\matS_a} 
\safemath{\subB}{\matS_b} 
\safemath{\dicta}{\matA} 	
\safemath{\dictb}{\matB} 	
\safemath{\hollowS}{H}
\safemath{\hollowA}{H_a}
\safemath{\hollowB}{H_b}
\safemath{\cross}{Z}
\safemath{\coh}{d}			
\safemath{\coha}{a}			
\safemath{\cohb}{b}			
\safemath{\dictset}{\setD}	
\safemath{\dictsetp}{\dictset(\coh,\coha,\cohb)}	
\safemath{\dictsetgen}{\dictset_\text{gen}}
\safemath{\dictsetgenp}{\dictsetgen(\coh)}
\safemath{\dictsetonb}{\dictset_\text{onb}}
\safemath{\dictsetonbp}{\dictsetonb(\coh)}

\safemath{\leftside}{U}
\safemath{\rightsideA}{R_a}
\safemath{\rightsideB}{R_b}

\safemath{\indexS}{\setI_S} 

\safemath{\na}{n_a}			
\safemath{\nb}{n_b}			
\safemath{\coeffa}{p_i}	
\safemath{\coeffb}{q_j}	
\safemath{\seta}{\setP}		
\safemath{\setb}{\setQ}     
\safemath{\setw}{\setW}	
\safemath{\setz}{\setZ}	
\safemath{\cola}{\veca}		
\safemath{\colb}{\vecb}		
\safemath{\cold}{\vecd}		
\safemath{\inputvec}{\vecx} 	
\safemath{\inputvecel}{x}
\safemath{\inputveca}{\vecx_a}
\safemath{\inputvecb}{\vecx_b}
\safemath{\outputvec}{\vecy}	
\safemath{\lambdamin}{\lambda_{\mathrm{min}}}
\DeclareMathOperator{\spark}{spark}
\newcommand{\pos}[1]{\lefto[#1\right]^+}
\newcommand{\normtwo}[1]{\vecnorm{#1}_2}
\newcommand{\normone}[1]{\vecnorm{#1}_1}
\newcommand{\normzero}[1]{\vecnorm{#1}_0}

\newcommand{\spectralnorm}[1]{\vecnorm{#1}}
\safemath{\elltwo}{\ell_2}
\safemath{\ellone}{\ell_1}
\safemath{\ellzero}{\ell_0}
\safemath{\ellinf}{\ell_\infty}
\safemath{\licardold}{Z(\dict)}
\safemath{\licard}{Z(\coh,\coha,\cohb)}
\safemath{\xsol}{\hat{x}}
\safemath{\xbord}{x_b}		
\safemath{\xstat}{x_s}		
\safemath{\xstatLone}{\tilde{x}_s}
\safemath{\order}{\mathcal{O}} 
\safemath{\orderlower}{\Omega}	
\safemath{\scales}{\Theta} 
\safemath{\ones}{\mathbf{1}} 
\safemath{\zeroes}{\mathbf{0}} 
\safemath{\thlone}{\kappa(\coh,\cohb)} 
\safemath{\constoneA}{\delta} 
\safemath{\constoneB}{\xi} 
\safemath{\nlarge}{L}				   
\safemath{\sumlarge}{S_\nlarge}
\safemath{\maxlarger}{P_\nlarge}	   
\safemath{\Pzero}{\textrm{P0}}	
\safemath{\Pone}{\textrm{P1}}
\safemath{\vecfir}{\vecw}			 
\safemath{\vecsec}{\vecz}
\safemath{\elvecfir}{w}              
\safemath{\elvecsec}{z}				 
\safemath{\nlargefir}{n}
\safemath{\normout}{\gamma}
\safemath{\auxfun}{h}
\safemath{\supp}{\textrm{supp}}
\safemath{\leqentry}{\stackrel{\trm{e}}{\leq}}
\safemath{\consta}{c_a}
\safemath{\constb}{c_b}
\safemath{\xmax}{x_{\text{max}}}

\begin{document}
	%
	%
\title{Uncertainty Relations and Sparse Signal Recovery for Pairs of General Signal Sets}

\author{Patrick Kuppinger,~\IEEEmembership{Student Member,~IEEE}, Giuseppe Durisi,~\IEEEmembership{Member,~IEEE}, and\\ Helmut B\"olcskei,~\IEEEmembership{Fellow,~IEEE}

\thanks{P. Kuppinger and H. B\"olcskei are with the Department of Information Technology and Electrical Engineering, ETH Zurich, Zurich, Switzerland, Email: \{patricku,boelcskei\}@nari.ee.ethz.ch}
\thanks{G. Durisi is with the Department of Signals and Systems, Chalmers University of Technology, Gothenburg, Sweden, Email: durisi@chalmers.se}
\thanks{The material in this paper was presented in part at the IEEE Information Theory Workshop (ITW), Taormina, Italy, October 2009, and at the IEEE International Symposium on Information Theory (ISIT), Austin, TX, June 2010.}
}

\maketitle
\begin{abstract}
We present an uncertainty relation for the representation of signals in two different general (possibly redundant or incomplete) signal sets. 
This uncertainty relation is relevant for the analysis of signals containing two distinct features each of which can be described sparsely in a suitable general signal set.
Furthermore, the new uncertainty relation is shown to lead to improved sparsity thresholds for recovery of signals that are sparse in general dictionaries.
%
%
Specifically, our results improve on the well-known $(1+1/\coh)/2$-threshold for dictionaries with coherence \coh by up to a factor of two.
Furthermore, we provide probabilistic recovery guarantees for pairs of general dictionaries that also allow us to understand which parts of a general dictionary one needs to randomize over to ``weed out'' the sparsity patterns that prohibit breaking the square-root bottleneck. 
%
        
%
\end{abstract}
\section{Introduction and Outline}\label{sec:intro}
%
%
A milestone in the sparse signal recovery literature is the uncertainty relation for the Fourier-identity pair found in~\cite{donoho1989}. This uncertainty relation was extended to pairs of arbitrary orthonormal bases (ONBs) in~\cite{elad2002}. Besides being interesting in their own right, these uncertainty relations are fundamental in the formulation of recovery guarantees for signals that contain two distinct features, each of which can be described sparsely using an ONB. If the individual features are, however, sparse only in overcomplete signal sets (i.e., in frames~\cite{christensen03}), the two-ONB result~\cite{donoho1989,elad2002} cannot be applied.
The goal of this paper is to find uncertainty relations and corresponding signal recovery guarantees for signals that are sparse in pairs of general (possibly redundant) signal sets.
Redundancy in the individual signal sets allows us to succinctly describe a wider class of features. 
Concrete examples for this setup can be found in the feature extraction or morphological component analysis literature (see, e.g., \cite{fadili-image,donoho2010} and references therein).
%
%

In order to put our results into perspective and to detail our contributions, we first briefly recapitulate the formal setup considered in the sparse signal recovery literature~\cite{donoho2001,donoho2002,elad2002,gribonval2003,tropp2004,candes2006b,tropp2008}.

\subsection{Sparse Signal Recovery Methods}\label{subsec:setup}
Consider the problem of recovering unknown vectors from small numbers of linear non-adaptive measurements.
More formally, let $\inputvec\in\complexset^\inputdim$ be an unknown vector that is observed through a measurement matrix \dict with columns\footnote{
Throughout the paper, we shall assume that the columns of \dict span $\complexset^{\outputdim}$ and have unit \elltwo-norm.} $\cold_i\in\complexset^\outputdim,\,\allo{i}{\inputdim}$, according to
\be
	\outputvec=\dict\inputvec
\een	
where $\outputvec\in\complexset^\outputdim$ and $\outputdim\ll\inputdim$. If we do not impose additional assumptions on \inputvec, the problem of recovering \inputvec from \outputvec is obviously ill-posed. The situation changes drastically if we assume that \inputvec is sparse in the sense of having only a few nonzero entries. More specifically, let $\normzero{\inputvec}$ denote the number of nonzero entries of \inputvec, then
\be
	(\Pzero)\quad \text{minimize } 
	\normzero{\inputvec}\quad\trm{subject to } \outputvec=\dict\inputvec
\een
can recover \inputvec without prior knowledge of the positions of the nonzero entries of \inputvec. 
%
%
Equivalently, we can interpret (\Pzero) as the problem of finding the sparsest representation of the vector \outputvec in terms of the ``dictionary elements'' (columns) $\cold_i$. In this context, the matrix \dict is often referred to as dictionary.

Since (\Pzero) is an NP-hard problem~\cite{davis1997} (it requires a combinatorial search), it is computationally infeasible, even for moderate problem sizes \inputdim, \outputdim.
%
Two popular and computationally more tractable alternatives to solving (\Pzero) are \emph{basis pursuit} (BP)~\cite{chen1998,donoho2001,donoho2002,gribonval2003,elad2002,tropp2004}
and \emph{orthogonal matching pursuit} (OMP)~\cite{Pati1993,davis1994,tropp2004}. 
BP is a convex relaxation of the (\Pzero) problem, namely
\be
	(\textrm{BP})\quad \text{minimize }
	\normone{\inputvec}\quad\trm{subject to } \outputvec=\dict\inputvec.
\een
Here, $\normone{\inputvec}=\sum_i \abs{x_i}$ denotes the \ellone-norm of the vector \inputvec.
OMP is an iterative greedy algorithm that constructs a sparse representation of \outputvec by selecting, in each iteration, the column of \dict most ``\emph{correlated}" with the difference between \outputvec and its current approximation.

Two questions that arise naturally are: \begin{inparaenum}[1)]
\item Under which conditions is \inputvec the unique solution of (\Pzero)?
\item Under which conditions is this solution delivered by BP and/or OMP?
\end{inparaenum}
Answers to these questions are typically expressed in terms of sparsity thresholds on the unknown vector \inputvec~\cite{donoho2001,donoho2002,gribonval2003,elad2002,tropp2004}.
These sparsity thresholds either hold for all possible sparsity patterns and values of nonzero entries in \inputvec, in which case we speak of \emph{deterministic} sparsity thresholds. 
Alternatively, one may be interested in so-called \emph{probabilistic} or---following the terminology used in~\cite{candes2006b}---\emph{robust} sparsity thresholds, which hold for  \emph{most} sparsity patterns and values of nonzero entries in \inputvec. Intuitively, robust sparsity thresholds are larger than deterministic ones. More precisely, as the number of measurements \outputdim grows large, deterministic sparsity thresholds generally scale at best as $\sqrt{\outputdim}$. Robust sparsity thresholds, in contrast, break this \emph{square-root bottleneck}. In particular, they scale on the order of $\outputdim/(\log\inputdim)$~\cite{tropp2008}.
However, this comes at a price: Uniqueness of the solution of\footnote{Whenever we speak of uniqueness of the solution of (\Pzero), we mean that the unique solution of (\Pzero) applied to $\outputvec=\dict\inputvec$ is given by \inputvec.} (\Pzero) and recoverability of the (\Pzero)-solution through BP is guaranteed only with high probability with respect to the choice of\footnote{Robust sparsity thresholds for OMP to deliver the unique (\Pzero)-solution are still unknown. For the \emph{multichannel} scenario, first results along these lines were reported in~\cite{gribonval2008atoms}, where it is shown that the probability of reconstruction error decays exponentially with the number of channels.} \inputvec.


%
%
%
Both deterministic and probabilistic sparsity thresholds are typically expressed in terms of the dictionary \emph{coherence}, defined as the maximum  absolute value over all inner products between
pairs of distinct columns of~\dict.

An alternative approach is to assume that the dictionary \dict is random (rather than the vector \inputvec) and to determine thresholds that hold for all (sufficiently) sparse \inputvec with high probability with respect to the choice of \dict~\cite{candes2006c,donoho2006,baraniuk2008a}. Throughout this paper, we consider deterministic dictionaries exclusively.

Note that when considering signals that consist of two distinct features, each of which can be described sparsely using an ONB~\cite{elad2002,donoho2001,feuer2003,tropp2004}, the corresponding dictionary \dict is given by the concatenation of these two ONBs. 
%
One obvious way of obtaining recovery guarantees for signals that are sparse in pairs of general signal sets is to \emph{concatenate} these general signal sets, view  the concatenation as one (general) dictionary, and apply the sparsity thresholds for general dictionaries reported in, e.g.,~\cite{donoho2002,gribonval2003,tropp2004,tropp2008}.
However, these sparsity thresholds depend only on the coherence of the resulting overall dictionary \dict and, in particular, do not take into account the coherence parameters of the two constituent signal sets.

%
%
In this paper, we show that the sparsity thresholds can be significantly improved not only if \dict is the concatenation of two ONBs---as was done in~\cite{elad2002,gribonval2003,feuer2003,tropp2004}---but also if \dict consists of the concatenation of two general signal sets (or \emph{sub-dictionaries}) with known coherence parameters.

\subsection{Contributions}
Our contributions can be detailed as follows.
Based on a novel uncertainty relation for pairs of general (redundant or incomplete) signal sets, we obtain a novel deterministic sparsity threshold guaranteeing (\Pzero)-uniqueness for dictionaries that are given by the concatenation of two general sub-dictionaries with known coherence parameters.
Additionally, we derive a novel  threshold guaranteeing that BP and OMP recover this unique (\Pzero)-solution.
Our thresholds improve significantly on the known deterministic sparsity thresholds one would obtain if the concatenation of two sub-dictionaries were viewed as a general dictionary, thereby ignoring the additional information about the sub-dictionaries' coherence parameters. 
More precisely, this improvement can be up to a factor of two.
Moreover, the known sparsity thresholds for general dictionaries and the ones for the concatenation of two ONBs follow from our results for the concatenation of general sub-dictionaries as special cases.

Concerning probabilistic sparsity thresholds for the concatenation of two general dictionaries, we address the following question: Given a \emph{general} dictionary, can we break the square-root bottleneck while only randomizing the sparsity patterns over a certain part of the overall dictionary?
By extending the known results for the two-ONB setting~\cite{candes2006b,tropp2008} to the concatenation of two general dictionaries, we show that the answer is in the affirmative. 
%
%
Our results 
allow us to identify parts of a general  dictionary the  sparsity patterns need to be randomized over so as to break the square-root bottleneck. 
\subsection{Notation}\label{subsec:nota}
We use lowercase boldface letters for column vectors, e.g., \vecx, and uppercase boldface letters for matrices, e.g., \dict. 
For a given matrix \dict, we denote its $i$th column by $\cold_i$, its conjugate transpose by $\herm{\dict}$, and its Moore-Penrose inverse by $\dict^{\dagger}$. 
Slightly abusing notation, we say that $\cold\in\dict$ if $\cold$ is a column of the matrix \dict.
The spectral norm of a matrix \dict is $\spectralnorm{\dict}=\sqrt{\lambda_\trm{max}(\dict^H\dict)}$, where $\lambda_\trm{max}(\dict^H\dict)$ denotes the maximum eigenvalue of $\dict^H\dict$. The minimum and maximum singular value of \dict are denoted by $\sigma_\text{min}(\dict)$ and $\sigma_\text{max}(\dict)$, respectively; 
$\rank(\dict)$ stands for the rank of \dict,  $\vecnorm{\dict}_{1,2}=\max_i\{\normtwo{\cold_i}\}$, and  $\vecnorm{\dict}_{1,1}=\max_i\{\normone{\cold_i}\}$. 
The smallest eigenvalue of the positive-semidefinite matrix $\matG$ is denoted by $\lambdamin(\matG)$. 
We use $\bI_n$ to refer to the $n\times n$ identity matrix; $\mathbf{0}_{m,n}$ and $\mathbf{1}_{m,n}$ stand for the all-zero and all-one matrix of size $m \times n$, respectively. We denote the $n$-dimensional  all-ones and all-zeros column vector by $\mathbf{1}_n$ and $\mathbf{0}_n$, respectively.
The natural logarithm is referred to as $\log$.
The set of all positive integers is $\naturals^+$.
For two functions $f(x)$ and $g(x)$, the notation $f(x)=\orderlower(g(x))$ means that there exists a real number $x_0$ such that $\abs{f(x)}\geq k_1\abs{g(x)}$ for all $x>x_0$, where $k_1$ is a finite constant. The notation $f(x)=\order(g(x))$ means that there exists a real number $x_0$ such that $\abs{f(x)}\leq k_2\abs{g(x)}$ for all $x>x_0$, where $k_2$ is a finite constant. Furthermore, we write $f(x)=\scales(g(x))$ if  there exists a real number $x_0$ and finite constants $k_1$ and $k_2$ such that $k_1\abs{g(x)}\leq\abs{f(x)}\leq k_2\abs{g(x)}$ for all $x>x_0$.
%
%
%
For $u\in\reals$,  we define $\pos{u}\!=\max\{0,u\}$.
Whenever we say that a vector $\inputvec\in\complexset^\inputdim$ has a \emph{randomly} chosen sparsity pattern of cardinality $L$, we mean that the support set of \inputvec (i.e., the set of nonzero entries of \inputvec) is chosen uniformly at random among all $\binom{\inputdim}{L}$ possible support sets of cardinality~$L$.

\section{Deterministic Sparsity Thresholds}\label{subsec:det}
\subsection{A Brief Review of Relevant Previous Work}\label{sec:det_rev}
%
%
%
A quantity that is intimately related to the uniqueness of the solution of (\Pzero)  is the \emph{spark} of a dictionary \dict, 
defined as the smallest number of linearly dependent columns of \dict~\cite{donoho2002}.
More specifically, the following result holds~\cite{donoho2002,gribonval2003}: For a given dictionary \dict and measurement outcome $\outputvec=\dict\inputvec$, the unique solution of (\Pzero) is given by \inputvec if
\be\label{eq:sparkP0}
	\normzero{\inputvec}<\frac{\spark(\dict)}{2}.
\ee
%
%
%
Unfortunately, determining the spark of a dictionary is an NP-hard problem, i.e., a problem that is as hard as solving (\Pzero) directly. It is possible, though, to derive easy-to-compute lower bounds on $\spark(\dict)$ that are explicit in the coherence of \dict defined as 
\be\label{eq:coherence}
	\coh = \max_{i\neq j}\abs{\herm{\cold_i}\cold_j}.
\ee
%
We next briefly review these lower bounds.
Let us first consider the case where \dict is the concatenation of two ONBs. Denote the set of all dictionaries that are the concatenation of two ONBs and have coherence \coh by \dictsetonbp.
It was shown in~\cite{elad2002} that for $\dict\in\dictsetonbp$, we have
\be\label{eq:EladSparkLower}
	\spark(\dict)\geq\frac{2}{\coh}.
\ee
Substituting~\eqref{eq:EladSparkLower} into~\eqref{eq:sparkP0} yields the following sparsity threshold guaranteeing that the unique solution of (\Pzero) applied to $\outputvec=\dict\inputvec$ is given by \inputvec:
\be\label{eq:ONB_P0}
	\normzero{\inputvec}<\frac{1}{\coh}.
\ee
Furthermore, it was shown in~\cite{elad2002,feuer2003,tropp2004} that for this unique solution to be recovered by BP and OMP it is sufficient to have
\be\label{eq:ONB_l1}
	\normzero{\inputvec}<\frac{\sqrt{2}-0.5}{\coh}\approx\frac{0.9}{\coh}.
\ee

A question that arises naturally is: What happens if the dictionary \dict is not the concatenation of two ONBs? There exist sparsity thresholds in terms of \coh for general dictionaries.
Specifically, let us denote the set of all dictionaries with coherence \coh by \dictsetgenp.
%
It was shown in~\cite{donoho2002,gribonval2003,tropp2004}  that for $\dict\in\dictsetgenp$ we have
\be\label{eq:spark_bound_donoho}
	\spark(\dict)\geq1+\frac{1}{\coh}.
\ee
Using~\eqref{eq:spark_bound_donoho} in~\eqref{eq:sparkP0} yields the following sparsity threshold guaranteeing that the unique solution of (\Pzero) applied to $\outputvec=\dict\inputvec$ is given by \inputvec: 
%
\be\label{eq:gen_sparsity_th}
	\normzero{\inputvec} < \frac{1}{2}\lefto(1+\frac{1}{\coh}\right).
\ee
Interestingly, one can show that~\eqref{eq:gen_sparsity_th} also guarantees that BP and OMP recover the unique (\Pzero)-solution~\cite{donoho2002,gribonval2003,tropp2004}.

The set \dictsetgenp is large, in general, and contains a variety of structurally very different dictionaries, ranging from equiangular tight frames (where the absolute values of the inner products between any two distinct dictionary elements are equal) to dictionaries where the maximum inner product is achieved by one pair only. 
The sparsity threshold in~\eqref{eq:gen_sparsity_th} is therefore inevitably rather crude.
Better sparsity thresholds are possible if one considers subsets of \dictsetgenp, such as, e.g.,  $\dictsetonbp\subset\dictsetgenp$.
A dictionary $\dict\in\dictsetonbp$ also satisfies $\dict\in\dictsetgenp$, and, hence, the sparsity threshold in~\eqref{eq:gen_sparsity_th} applies. However, the additional structural information about \dict being the concatenation of two ONBs, i.e., $\dict\in\dictsetonbp$, allows us to obtain the improved sparsity thresholds in~\eqref{eq:ONB_P0} and~\eqref{eq:ONB_l1}, which are (for $\coh\ll1$) almost a factor of two higher (better) than the threshold in~\eqref{eq:gen_sparsity_th}.
As a side remark, we note that the threshold for the two-ONB case in~\eqref{eq:ONB_l1} drops below that in~\eqref{eq:gen_sparsity_th}, valid for general dictionaries, if $\coh>2(\sqrt{2}-1)$. This is surprising as exploiting structural information should lead to a higher sparsity threshold. We will show, in~\fref{sec:det_new}, that one can refine the threshold in~\eqref{eq:ONB_l1} so as to fix this problem.  

\subsection{Novel Deterministic Sparsity Thresholds for the Concatenation of Two General Signal Sets}\label{sec:det_new}
We consider dictionaries with coherence \coh that consist of two  sub-dictionaries with coherence \coha and \cohb, respectively. 
The set of all such dictionaries will be denoted as \dictsetp. 
A dictionary $\dict\in\dictsetp$ of dimension $\outputdim\times\inputdim$ (with $\inputdim\geq\outputdim$) can be written as $\dict=[\dicta\,\,\dictb]$, where the sub-dictionary $\dicta\in\complexset^{\outputdim\times\inputdimA}$ has coherence \coha and the sub-dictionary $\dictb\in\complexset^{\outputdim\times\inputdimB}$ has  coherence \cohb.
We remark that the two sub-dictionaries need not be ONBs, need not have the same number of elements and need not span $\complexset^\outputdim$, but their concatenation is assumed to span $\complexset^\outputdim$. Without loss of generality, we assume, throughout the paper, that $\coha\leq\cohb$.
For fixed \coh,\footnote{We assume throughout the paper that $\coh>0$. For $\coh=0$ the dictionary \dict consists of orthonormal columns, and, hence, every unknown vector \inputvec can be uniquely recovered from the measurement outcome \outputvec according to $\inputvec=\dict^H\outputvec$.} we have that $\dictsetp\subset\dictsetgenp$.
Hence, we consider  subsets \dictsetp of the set \dictsetgenp parametrized by the coherence parameters \coha and \cohb. 
%

%
%
For $\dict\in\dictsetp$ we derive sparsity thresholds in terms of \coh, \coha, and \cohb and show that these thresholds improve upon those in~\eqref{eq:gen_sparsity_th} for general dictionaries $\dict\in\dictsetgenp$.
This improvement is a result of the restriction to a subset of dictionaries in \dictsetgenp, namely \dictsetp, and of exploiting the  additional structural information (in terms of the coherence parameters \coha and \cohb) available about dictionaries \dict in this subset.
%
%
%
%

Every dictionary in \dictsetgenp can be viewed as the concatenation of two sub-dictionaries. Our results therefore state that viewing a dictionary $\dict\in\dictsetgenp$ as the concatenation of two sub-dictionaries 
leads to improved sparsity thresholds provided that the coherence parameters \coha and \cohb of the respective sub-dictionaries are available.
Moreover, the improvements will be seen to be up to a factor of two if \coha and \cohb are sufficiently small.
%
%

The sparsity threshold for uniqueness of the solution of (\Pzero) for dictionaries $\dict\in\dictsetp$, formalized in~\fref{thm:spark_bound} below, is based on a novel uncertainty relation for pairs of general dictionaries, stated in the following lemma. 
\begin{lem}\label{lem:uncertainty}
	Let $\dicta\in\complexset^{\outputdim\times\inputdimA}$ be a dictionary with coherence \coha,  $\dictb\in\complexset^{\outputdim\times\inputdimB}$ a dictionary with coherence \cohb, and denote the coherence of the concatenated dictionary $\dict=[\dicta\,\,\dictb]$, $\dict \in \complexset^{\outputdim \times \inputdim}$, by \coh. 
	For every vector $\vecs\in\complexset^\outputdim$
 that can be represented as a linear combination of~$\na$ columns of~\dicta and, equivalently, as a linear combination of~$\nb$  columns of~\dictb,\footnote{For $\na=0$ or $\nb=0$ we define $\vecs=\mathbf{0}_\outputdim$. We exclude the trivial case $\na=\nb=0$.
 }  the following inequality holds:
	\be
	\label{eq:uncertainty}
		\na\nb\geq \dfrac{\pos{1-\coha(\na-1)}\pos{1-\cohb(\nb-1)}}{\coh^2}.
	\ee
\end{lem}
\begin{IEEEproof}
See~\fref{app:uncertainty_lemma}.
\end{IEEEproof}
%
The uncertainty relation for the union of two-ONB case derived in~\cite{elad2002} is a special case of~\eqref{eq:uncertainty}. In particular,  if $\coha=\cohb=0$, then~\eqref{eq:uncertainty} reduces to the result reported in~\cite[Thm. 1]{elad2002}: 
\be\label{eq:uncertaintyONB}
	\elemA\elemB\geq\frac{1}{\coh^2}.
\ee
Note that, differently from~\cite[Thm. 1]{elad2002},  the lower bound in~\eqref{eq:uncertaintyONB} holds not only for the concatenation of two ONBs, but also for the concatenation of two sub-dictionaries \dicta and \dictb that contain orthonormal columns but individually do not necessarily span $\complexset^\outputdim$ (but their concatenation spans $\complexset^\outputdim$).
\fref{lem:uncertainty} allows us to easily recover several other well-known results 
such as, e.g., the well-known lower bound in~\eqref{eq:spark_bound_donoho} on the spark of a dictionary.
To see this note that when $\nb=0$ in~\fref{lem:uncertainty} (and thus $\vecs=\mathbf{0}_\outputdim$, by definition) then the \na columns in \dicta participating in the representation of \vecs are linearly dependent. 
%
Moreover, for $\nb=0$ we have $\pos{1-\cohb(\nb-1)}=(1+\cohb)>0$. Therefore, it follows from~\eqref{eq:uncertainty} that necessarily $\pos{1-\coha(\na-1)}=0$ and thus $\na\geq1+1/\coha$, which agrees with the lower bound on the spark of the (sub-)dictionary \dicta~\cite{donoho2002,gribonval2003,tropp2004}. 
A similar observation follows for $\na=0$.

More importantly, \fref{lem:uncertainty} also allows us to derive a new lower bound on the spark of the overall dictionary $\dict=[\dicta\,\,\dictb]\in\dictsetp$.
When used in~\eqref{eq:sparkP0}, this result then yields a new sparsity threshold guaranteeing uniqueness of the (\Pzero)-solution. We show that this threshold improves upon that in~\eqref{eq:gen_sparsity_th}, which would be obtained if we viewed \dict simply as a general dictionary in \dictsetgenp, thereby ignoring the fact that the dictionary under consideration belongs to a subset, namely \dictsetp, of \dictsetgenp.

\begin{thm}\label{thm:spark_bound}
	For  $\dict\in\dictsetp$, a sufficient condition for the vector \inputvec to be the unique solution of (\Pzero) applied to $\outputvec=\dict\inputvec$ is that
\be\label{eq:P0bound}
	\normzero{\inputvec} < \frac{f(\xsol)+\xsol}{2}
\ee
%
	%
	where
	\be
	f(x)= \dfrac{(1+\coha)(1+\cohb)-x\cohb(1+\coha)}{x(\coh^2-\coha\cohb)+\coha(1+\cohb)}
	\een
	and
	%
		$\xsol=\min\{\xbord,\xstat\}$.
	%
	Furthermore,
	\be
		\xbord=\dfrac{1+\cohb}{\cohb+\coh^2}
	\een
	and
	\be
	\xstat=\begin{cases}
			\dfrac{1}{\coh}, & \text{if}\quad \coha=\cohb=\coh, \\[5mm]
			\dfrac{\coh\sqrt{(1+\coha)(1+\cohb)}-\coha-\coha\cohb}{\coh^2-\coha\cohb}, &\text{otherwise.}
		   \end{cases}		
	\een	

\end{thm}
\begin{IEEEproof}
		See \fref{app:spark_bound}. 
\end{IEEEproof}
The sparsity threshold in~\eqref{eq:P0bound} reduces to that in~\eqref{eq:gen_sparsity_th} when~$\cohb=\coh$ (irrespective of \coha) or when $\coh=1$ (irrespective of \coha and \cohb). 
Hence, the sparsity threshold in~\eqref{eq:P0bound} does not improve upon that in~\eqref{eq:gen_sparsity_th} if the pair of columns achieving the overall dictionary coherence \coh appears in the same sub-dictionary \dictb (recall that we assumed $\cohb\geq\coha$), or if $\coh=1$.
%
In all other cases, the sparsity threshold in~\eqref{eq:P0bound} can be shown to be strictly larger than that in~\eqref{eq:gen_sparsity_th}. 
This result is proven in \fref{app:betterSparkbound}.
The improvement can be up to a factor of two. We demonstrate this for the special case $\coha=\cohb$, for which the sparsity threshold in~\eqref{eq:P0bound} takes a particularly simple form.
%
In this case, as can easily be verified, $\xstat\leq\xbord$ so that~\eqref{eq:P0bound} reduces to
\be\label{eq:P0symmetric}
	\normzero{\inputvec}<\dfrac{1+\cohb}{\coh+\cohb}.
\ee

\begin{figure}[t]
\vspace{2mm}
\begin{center}
 \includegraphics[width=\linewidth]{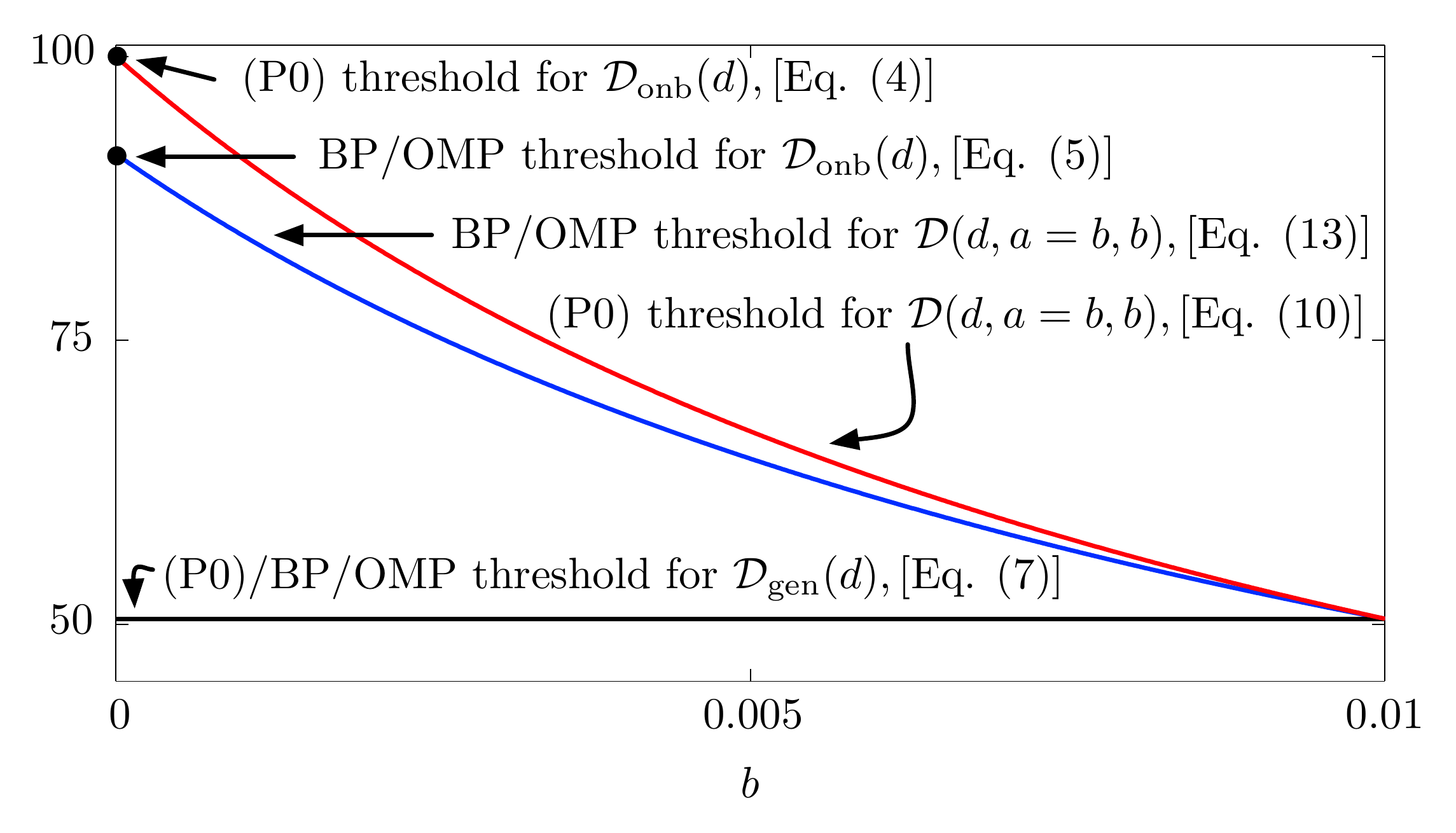}
  \caption{Deterministic sparsity thresholds guaranteeing uniqueness of (\Pzero) and recoverability via BP and OMP for dictionaries in~\dictsetgenp, \dictsetonbp, and \dictsetp. We set $\coh=0.01$ and consider the special case $\coha=\cohb$. Note that for $\coha=\cohb$, the threshold in~\eqref{eq:P0bound} reduces to that in~\eqref{eq:P0symmetric}.}
   \label{fig:detThreshold}
\end{center}
\end{figure}

For $\coha=\cohb=0$ the sparsity threshold in~\eqref{eq:P0symmetric} reduces to the known sparsity  threshold for dictionaries in \dictsetonbp specified in~\eqref{eq:ONB_P0}. 
Note, however, that the threshold in~\eqref{eq:P0symmetric} with $\cohb=0$ holds for all $\dict\in\dictset(0,0,\coh)$, thereby also including  sub-dictionaries \dicta and \dictb that contain orthonormal columns but do not necessarily individually span $\complexset^\outputdim$ (but their concatenation spans $\complexset^\outputdim$).
%
%
Setting $\cohb=\epsilon\coh$ with $\epsilon\in[0,1]$ and noting that for $\coh\ll 1$ the ratio between the sparsity threshold in~\eqref{eq:P0symmetric} and that in~\eqref{eq:gen_sparsity_th} is roughly $2/(1+\epsilon)$, which for $\epsilon\ll1$ is almost  two. Note that, for small coherence parameters \coha and \cohb, the elements in each of the two sub-dictionaries \dicta and \dictb are close to being orthogonal to each other. 
\fref{fig:detThreshold} shows the sparsity threshold in~\eqref{eq:P0symmetric} for $\coh=0.01$ as a function of \cohb. We can see that for $\cohb\ll\coh$ the threshold in~\eqref{eq:P0symmetric} is, indeed, almost a factor of two larger than that in~\eqref{eq:gen_sparsity_th}.

So far, we focused on thresholds guaranteeing (\Pzero)-uniqueness.
We next present thresholds guaranteeing recovery of the unique (\Pzero)-solution via BP and OMP for dictionaries $\dict\in\dictsetp$.
The recovery conditions we report in~\fref{thm:l1_na_nb} and~\fref{cor:l1_general} below, depend on \cohb and \coh, but not on \coha. 
Slightly improved thresholds that also depend on \coha can be derived following similar ideas as in the proofs of~\fref{thm:l1_na_nb} and~\fref{cor:l1_general}.
The resulting expressions are, however, unwieldy and will therefore not be presented here. 
\begin{thm}\label{thm:l1_na_nb}
	Suppose that ~$\outputvec \in\complexset^\outputdim$ can be represented as $\outputvec=\dict\inputvec$, where \inputvec has \elemA nonzero entries corresponding to columns of \dicta and \elemB nonzero entries corresponding to columns of \dictb. Without loss of generality, we assume that $\elemA\leq \elemB$.
	A sufficient condition for BP \emph{and} OMP to recover \inputvec is
	\begin{align}
		\label{eq:BP_OMP_recovery_na_nb}
		2\elemA(1+\cohb)\cohb+\elemB(1+\cohb)(\coh+\cohb)+2\elemA\elemB(\coh^2-\cohb^2)<(1+\cohb)^2.
	\end{align}
\end{thm}
\begin{IEEEproof}
See~\fref{app:l1proof}.
\end{IEEEproof}
\fref{thm:l1_na_nb} generalizes the result in \cite[Sec. 6]{elad2002}, \cite[Cor. 3.8]{tropp2004} for the concatenation of two ONBs to dictionaries $\dict\in\dictsetp$. 
In particular,~\eqref{eq:BP_OMP_recovery_na_nb} reduces to \cite[Eq. (16)]{tropp2004} when $\cohb=0$ (since $\coha\leq\cohb$, this implies $\coha=0$).
Furthermore, when $\cohb=\coh$, the condition in~\eqref{eq:BP_OMP_recovery_na_nb} simplifies to $\na+\nb<(1+1/\coh)/2$, thereby recovering the sparsity threshold in~\eqref{eq:gen_sparsity_th}. Thus, if the pair of columns achieving the overall dictionary coherence is in the same sub-dictionary \dictb (recall that we assumed $\cohb\geq\coha$),  no improvement over the well-known $(1+1/\coh)/2$-threshold for dictionaries in \dictsetgenp is obtained.
\fref{thm:l1_na_nb} depends explicitly  on \na and \nb. In the following corollary, we provide a recovery guarantee in the form of a sparsity threshold that depends on \na and \nb only through the overall sparsity level of \inputvec according to $\normzero{\inputvec}=\na+\nb$.
\begin{cor}\label{cor:l1_general}
For  $\dict\in \dictsetp$ a sufficient condition for BP \emph{and} OMP to deliver the unique solution of (\Pzero) is 
\be
	\label{eq:l1_final}
		\normzero{\inputvec}< 
		\begin{cases}
	\dfrac{(1+\cohb)\bigl[\constoneB-(\coh+3\cohb)\bigr]}{2(\coh^2-\cohb^2)},& \text{if } \cohb<\coh  \text{ and }  \thlone >1,\\[5mm]
		\dfrac{1+2\coh^2 +3\cohb -\coh(1+\cohb)}{2(\coh^2+\cohb)}, &\text{otherwise}
		\end{cases}
	\ee
	%
	with
	\be
	\label{eq:condition_corollary_omp_bp}
		\thlone=\dfrac{(1+\cohb)(\constoneB-4\cohb)}{4(\coh^2-\cohb^2)}
	\ee
and $\constoneB=2\sqrt{2}\sqrt{\coh(\cohb+\coh)}$.
\end{cor} 
\begin{IEEEproof}
See~\fref{app:proof_cor_BP_OMP}.
\end{IEEEproof}
The sparsity threshold in~\eqref{eq:l1_final} reduces to the sparsity threshold in~\eqref{eq:gen_sparsity_th} when $\cohb=\coh$ or when $\coh=1$ (irrespective of \cohb). 
%
%
In all other cases, the sparsity threshold in~\eqref{eq:l1_final} is strictly larger than that in~\eqref{eq:gen_sparsity_th} (see~\fref{app:proofl1better}). 
The threshold in~\eqref{eq:l1_final} is complicated as we have to deal with two different cases.
The distinction between these two cases is, however, crucial to ensure that the threshold in~\eqref{eq:l1_final} does not fall below that in~\eqref{eq:gen_sparsity_th}.\footnote{Recall that for $\coh>2(\sqrt{2}-1)$ the threshold in~\eqref{eq:ONB_l1} drops below that in~\eqref{eq:gen_sparsity_th}.}
%
It turns out that the first case in~\eqref{eq:l1_final} is active whenever $\cohb<\coh<3/5$, which covers essentially all practically relevant cases. In fact, for dictionaries with coherence $\coh\geq3/5$, the sparsity threshold in~\eqref{eq:l1_final} allows for at most one nonzero entry in \inputvec.

The improvement of the sparsity threshold in~\eqref{eq:l1_final} over that in~\eqref{eq:gen_sparsity_th} can be up to a factor of almost two. This can be seen by setting $\cohb=\epsilon\coh$ with $\epsilon\in[0,1)$ and noting that for $\coh\ll1$ the ratio between the sparsity threshold in the first case in~\eqref{eq:l1_final} and that in~\eqref{eq:gen_sparsity_th}  is roughly $(2\sqrt{2(1+\epsilon)}-(1+3\epsilon))/(1-\epsilon^2)$, which for $\epsilon\ll1$ is approximately $1.8$.
\fref{fig:detThreshold} shows the threshold in~\eqref{eq:l1_final} for $\coh=0.01$ as a function of \cohb. We can see that for $\cohb\ll\coh$ the threshold in~\eqref{eq:l1_final} is, indeed, almost a factor of two larger than that in~\eqref{eq:gen_sparsity_th}.

If \dict is the concatenation of two ONBs, and hence $\coha=\cohb=0$,  the sparsity threshold in~\eqref{eq:l1_final} reduces~to
\be
\label{eq:new_l1_orth}
		\normzero{\inputvec} < 
		\begin{cases}\dfrac{\sqrt{2}-0.5}{\coh},& \text{if } \coh<\dfrac{1}{\sqrt{2}},\\[5mm]
		1+\dfrac{1-\coh}{2\coh^2},&\text{otherwise}.
		\end{cases}
\ee
For $\coh<1/\sqrt{2}$, this threshold is the same as that in~\eqref{eq:ONB_l1} but improves on~\eqref{eq:ONB_l1} if $\coh\geq1/\sqrt{2}$. In particular, unlike the threshold in~\eqref{eq:ONB_l1}, 
the threshold in~\eqref{eq:new_l1_orth} is guaranteed to be at least as large as that in~\eqref{eq:gen_sparsity_th}.

\section{Robust Sparsity Thresholds}\label{sec:prob}
The deterministic sparsity thresholds for dictionaries in \dictsetp derived in the previous section (as those available in the literature for dictionaries in \dictsetonbp and \dictsetgenp) all suffer from the so-called square-root bottleneck~\cite{tropp2008}.
Specifically, from the Welch lower bound on coherence~\cite{welch1974lower}
 \be
 	\coh\geq\sqrt{\frac{\inputdim-\outputdim}{\outputdim(\inputdim-1)}}
 \een
we can conclude that, for $\inputdim\gg\outputdim$, the deterministic sparsity thresholds  reported in this paper scale as $\sqrt{\outputdim}$ as \outputdim grows large. Put differently, for a fixed number of nonzero entries \sparsity in \inputvec, i.e., for a fixed sparsity level, the number of measurements \outputdim required to recover \inputvec through (\Pzero), BP, or OMP is on the order of $\sparsity^2$. The square-root bottleneck stems from the fact that deterministic sparsity thresholds are universal thresholds in the sense of applying to all  possible sparsity patterns (of cardinality \sparsity) and values of the corresponding nonzero entries of \inputvec. 
%
%
As already mentioned in~\fref{sec:intro}, the probabilistic (i.e., robust) sparsity thresholds scale fundamentally better, namely according to $\outputdim/\log\inputdim$, which implies that the number of measurements required to recover \inputvec is on the order of $\sparsity\log\inputdim$ instead of $\sparsity^2$.
%
%
%

%

%
We next address the following question: Given a \emph{general} dictionary, can we break the square-root bottleneck by only randomizing the sparsity patterns over a certain part of the overall dictionary? The answer turns out to be in the affirmative. 
It was shown in~\cite{candes2006c,tropp2008}---for the concatenation of two ONBs---that randomization of the sparsity patterns is only required over one of the two ONBs.
Before stating our results for general dictionaries let us briefly summarize the known results for concatenations of ONBs.


\subsection{A Brief Review of Relevant Previous Work}\label{sec:prob_rev}
%

%
Robust sparsity thresholds for the concatenation of two ONBs were first reported in~\cite{candes2006b} (based on earlier work in~\cite{candes2006c}) and later improved in~\cite{tropp2008}. 
In~\fref{thm:Tropp_ONB} below, we restate a result from~\cite{tropp2008} (obtained by combining Theorems D, 13, and 14) in a slightly modified form better suited to draw parallels to the  case of dictionaries in \dictsetp considered in this paper.
\begin{thm}[Tropp, 2008]\label{thm:Tropp_ONB}
Assume that\footnote{In~\cite{tropp2008} it is assumed that $\outputdim\geq3$ (and hence $\inputdim\geq6$). However, it can be shown that $\inputdim>2$ is sufficient to establish the result.} $\inputdim>2$. Let $\dict\in\complexset^{\outputdim\times\inputdim}$ be the union of two ONBs for $\complexset^\outputdim$ given by \dicta and \dictb  (i.e., $\inputdim=2\outputdim$) and denote the coherence of \dict as \coh. Fix $s\geq1$. Let the vector $\inputvec\in\complexset^\inputdim$ have an \emph{arbitrarily} chosen sparsity pattern of \elemA nonzero entries corresponding to columns of sub-dictionary \dicta and a \emph{randomly} chosen sparsity pattern of \elemB nonzero entries corresponding to columns of sub-dictionary \dictb. Suppose that 
\be\label{eq:Tropp1}
	\elemA+\elemB<\min\lefto\{\frac{c\,\coh^{-2}}{s\log\inputdim},\frac{\coh^{-2}}{2}\right\}
\ee
%
where $c=0.004212$. 
If the entries of \inputvec restricted to the chosen sparsity pattern are jointly continuous random variables,\footnote{For a definition of joint continuity, we refer to~\cite[pp.~40]{grimmett}.} then the unique solution of (\Pzero) applied to $\outputvec=\dict\inputvec$ is given by \inputvec with probability exceeding $(1-\inputdim^{-s})$.

If  the total number of nonzero entries satisfies
\be\label{eq:Tropp2}
	\elemA+\elemB<\min\lefto\{\frac{c\,\coh^{-2}}{s\log\inputdim},\frac{\coh^{-2}}{2},\frac{\coh^{-2}}{8(s+1)\log\inputdim}\right\}
\ee
and the entries of \inputvec restricted to the chosen sparsity pattern are jointly continuous random variables with i.i.d. phases that are uniformly distributed in $[0,2\pi)$ (the magnitudes need not be i.i.d.), then the unique solution of both (\Pzero) and BP applied to $\outputvec=\dict\inputvec$ is given by \inputvec with probability exceeding $(1-3\inputdim^{-s})$.
\end{thm}
\vspace{3mm}

An important consequence of~\fref{thm:Tropp_ONB} is the following: For the concatenation of two ONBs a robust sparsity threshold $\sparsity=\na+\nb$ of order $\outputdim/(\log\inputdim)$ is possible if the coherence \coh of the overall dictionary is on the order of $1/\sqrt{\outputdim}$. 
Note that for the same coherence \coh, deterministic sparsity thresholds would suffer from the square-root bottleneck as discussed in~\cite{tropp2008}.
%
%
Remarkably,~\fref{thm:Tropp_ONB} does not require that  the positions of \emph{all} nonzero entries of \inputvec are chosen randomly: It suffices to pick the positions of the nonzero entries of \inputvec corresponding to one of the two ONBs at random, while the positions of the remaining nonzero entries---all corresponding to columns in the other ONB---can be chosen arbitrarily. 
This essentially means that the result is universal with respect to one of the two ONBs (\dicta by choice of notation here) in the sense that \emph{all} possible combinations of \na columns in \dicta are allowed. Randomization over the other ONB ensures that the overall sparsity patterns that cannot be recovered (with on the order of $\sparsity\log\inputdim$ measurements) are ``weeded out''.
Moreover, randomization is needed on the values of \emph{all} nonzero entries of \inputvec, which reflects the fact that there exist certain value assignments on a given sparsity pattern that cannot be recovered with on the order of  $\sparsity\log\inputdim$ measurements.
In summary,~\fref{thm:Tropp_ONB} states that every sparsity pattern in \dicta in conjunction with most sparsity patterns in \dictb and most value assignments on the resulting overall sparsity pattern can be recovered. 

This result is interesting as it hints at the possibility of isolating specific parts of the dictionary \dict that require randomization to ``weed out'' the support sets that are not recoverable.
Unfortunately, the two-ONB structure is too restrictive to bring out this aspect.
Specifically, as the two ONBs are on equal footing, the result in~\fref{thm:Tropp_ONB} does not allow us to understand which properties of a sub-dictionary are responsible for problematic sparsity patterns. 
This motivates looking at robust sparsity thresholds for the concatenation of two general dictionaries.
Now, we could interpret the concatenation of two general (sub-)dictionaries as a general dictionary in \dictsetgenp and apply the robust sparsity thresholds for general dictionaries reported in~\cite{tropp2008}. This  requires, however, randomization over the entire dictionary (i.e., the positions of all nonzero entries of \inputvec have to be chosen at random and the values as well).
Hence, the robust sparsity threshold for general dictionaries does not allow us to isolate specific parts of the dictionary \dict that require randomization to ``weed out'' the support sets that are not recoverable with on the order of  $\sparsity\log\inputdim$ measurements.

\subsection{Robust Sparsity Thresholds for the Concatenation of General Signal Sets}\label{sec:prob_new}
We next derive robust sparsity thresholds for dictionaries $\dict\in\dictsetp$. Our results not only generalize~\fref{thm:Tropp_ONB} to the concatenation of two general dictionaries but, since every dictionary in \dictsetgenp can be viewed as the concatenation of two sub-dictionaries, also allow us to understand which part of a general dictionary requires randomization to ``weed out'' the support sets that are not recoverable.
%
%
%
\begin{thm}\label{thm:l0_l1_unique}
Assume that $\inputdim>2$. Let $\dict=[\dicta\,\,\dictb]$ be a dictionary in \dictsetp. Fix $s\geq1$ and $\gamma\in[0,1]$. 
Consider a random vector $\inputvec=\lefto[\inputveca^T\,\,\inputvecb^T\right]^T$ where $\inputveca\in\complexset^\inputdimA$ has an \emph{arbitrarily} chosen sparsity pattern of cardinality \elemA such that
\be\label{eq:condA}
	6\sqrt{2}\sqrt{\elemA\coh^2s\log\inputdim}+2(\elemA-1)\coha\leq(1-\gamma) e^{-1/4}
\ee
and $\inputvecb\in\complexset^\inputdimB$ has a \emph{randomly} chosen sparsity pattern of cardinality\footnote{Since we will be interested in the individual scaling behavior of \na and \nb as \outputdim grows large, we shall assume in the remainder of the paper that $\na,\nb\geq1$.} \elemB such that 
%
\be\label{eq:condB}
	24\sqrt{\elemB\cohb^2s\log\inputdim}+\frac{4\elemB}{\inputdimB}\spectralnorm{\dictb}^2+2\sqrt{\frac{\elemB}{\inputdimB}}\spectralnorm{\dicta}\!\spectralnorm{\dictb}\leq\gamma e^{-1/4}.
\ee
%

If the total number of nonzero entries of \inputvec satisfies
\be\label{eq:cond_l0}
	\elemA+\elemB\leq\frac{\coh^{-2}}{2}
\ee
and the entries of \inputvec restricted to the chosen sparsity pattern are jointly continuous random variables, then the unique solution of (\Pzero) applied to $\outputvec=\dict\inputvec$ is given by \inputvec with probability exceeding $(1-\inputdim^{-s})$. 

If the total number of nonzero entries of \inputvec satisfies
\be\label{eq:cond_l1}
		\elemA+\elemB<\min\left\{\frac{\coh^{-2}}{2},\frac{\coh^{-2}}{8(s+1)\log\inputdim}\right\}
\ee
and the entries of \inputvec restricted to the chosen sparsity pattern are jointly continuous random variables with  i.i.d. phases that are uniformly distributed in $[0,2\pi)$ (the magnitudes need not be i.i.d.),
then the unique solution of both (\Pzero) and BP applied to $\outputvec=\dict\inputvec$ is given by \inputvec with probability exceeding $(1-3\inputdim^{-s})$.
\end{thm}
\vspace{3mm}
\begin{IEEEproof}
The proof is based on the following lemma proven in \fref{app:proofs}.
\begin{lem}\label{lem:subdictionary}
Fix $s\geq1$ and $\gamma\in[0,1]$. Let \subD be a sub-dictionary of $\dict=[\dicta\,\,\dictb]\in\dictsetp$ containing \elemA  \emph{arbitrarily} chosen columns of \dicta and \elemB \emph{randomly} chosen columns of \dictb. If \elemA and \elemB satisfy~\eqref{eq:condA} and~\eqref{eq:condB}, respectively, then the minimum singular value $\sigma_\text{min}(\subD)$ of the sub-dictionary \subD obeys
\be
	\Prob\lefto\{\sigma_\text{min}(\subD)\leq \frac{1}{\sqrt{2}}\right\}\leq \inputdim^{-s}.
\een
\end{lem}
\vspace{3mm}

The proof of~\fref{thm:l0_l1_unique} then follows from~\fref{lem:subdictionary} and the results in~\cite{tropp2008} as follows.
The sparsity pattern of \inputvec obtained according to the conditions in~\fref{thm:l0_l1_unique} induces a sub-dictionary \subD of \dict containing \elemA arbitrarily chosen columns of \dicta and \elemB randomly chosen columns of \dictb. As a consequence of~\fref{lem:subdictionary}, the smallest singular value of this sub-dictionary exceeds $1/\sqrt{2}$ with probability at least $(1-\inputdim^{-s})$. 

\fref{lem:subdictionary} together with condition~\eqref{eq:cond_l0} and the requirement that the entries of \inputvec restricted to the chosen sparsity pattern are jointly continuous random variables implies, as a consequence of~\cite[Thm. 13]{tropp2008} (see also~\fref{app:Troppresults} where~\cite[Thm. 13]{tropp2008} is restated for completeness), that  the unique solution of (\Pzero) applied to $\outputvec=\dict\inputvec$ is given by \inputvec with probability at least $(1-\inputdim^{-s})$.

The second statement in~\fref{thm:l0_l1_unique} is proven as follows. \fref{lem:subdictionary}, together with condition~\eqref{eq:cond_l1}, and the requirement that the entries of \inputvec restricted to the chosen sparsity pattern are jointly continuous random variables with i.i.d. phases that are uniformly distributed in $[0,2\pi)$, implies, as a consequence of~\cite[Thm. 13]{tropp2008} and~\cite[Thm. 14]{tropp2008} (see also~\fref{app:Troppresults}), that the unique solution of \emph{both} (\Pzero) \emph{and} BP applied to $\outputvec=\dict\inputvec$ is given by \inputvec with probability at least $(1-\inputdim^{-s})(1-2\inputdim^{-s})\geq(1-3\inputdim^{-s})$.
\end{IEEEproof}
\vspace{0.2cm}

~\fref{thm:l0_l1_unique} generalizes the result in~\fref{thm:Tropp_ONB} to the concatenation $\dict=[\dicta\,\,\dictb]$ of the general dictionaries \dicta and \dictb. 
Next, we determine conditions on $\dict=[\dicta\,\,\dictb]$ for breaking the square-root bottleneck. 
More precisely, we determine conditions on $\dict=[\dicta\,\,\dictb]$ such that for vectors \inputvec with\footnote{Whenever for some function $g(\outputdim,\inputdim)$ we write $\scales(g(\outputdim,\inputdim))$, $\orderlower(g(\outputdim,\inputdim))$, or $\order(g(\outputdim,\inputdim))$, we mean that the ratio $\outputdim/\inputdim$ remains fixed while $\outputdim\to\infty$.} $\elemA=\scales(\outputdim/(\log\inputdim))$ and $\elemB=\scales(\outputdim/(\log\inputdim))$ the unique solution of \emph{both} (\Pzero) \emph{and} BP applied to $\outputvec=\dict\inputvec$ is given by \inputvec with probability at least $1-3\inputdim^{-s}$. This implies a robust sparsity threshold $\sparsity=\elemA+\elemB$ of  $\scales(\outputdim/(\log\inputdim))$. Note that we say the square-root bottleneck is broken only if both \na and \nb are on the order of $\outputdim/(\log\inputdim)$.

Conditions~\eqref{eq:condA}--\eqref{eq:cond_l1} in~\fref{thm:l0_l1_unique} yield upper bounds on the possible values of \na and \nb (such that the unique solution of both (\Pzero) and BP is given by \inputvec) that depend on the dictionary parameters \coh, \coha, \cohb, \inputdimA, \inputdimB, and the spectral norms of \dicta and \dictb. In the following, we rewrite these upper bounds by absorbing all constants (including $\gamma$ and $s$ defined in~\fref{thm:l0_l1_unique}) that are independent of   \coh, \coha, \cohb, \inputdimA, \inputdimB, and the spectral norms of \dicta and \dictb in a constant $c$. Note that $c$ can take on a different value at each appearance. 
We then derive necessary and sufficient conditions on the dictionary parameters \coh, \coha, \cohb, \inputdimA, \inputdimB, and the spectral norms of \dicta and \dictb for the resulting upper bounds on \na and \nb to be on the order of $\outputdim/(\log\inputdim)$, respectively.
%
%
%

We start with condition~\eqref{eq:condA}, which, together with the obvious condition $\elemA\leq\inputdimA$, yields the following constraint on \elemA:
\be
	\elemA\leq c\min\lefto\{\frac{\coh^{-2}}{\log\inputdim},\coha^{-1},\inputdimA\right\}.
\een
As $\outputdim$ grows large, this upper bound is compatible\footnote{We say that an upper bound on \na, \nb is compatible with  the scaling behavior $\scales(\outputdim/(\log\inputdim))$, if it does not preclude this scaling behavior.} 
 with the scaling behavior $\elemA=\scales(\outputdim/(\log\inputdim))$ if and only if all of the following conditions are met:
\renewcommand{\labelenumi}{\roman{enumi})}
\begin{enumerate}
	\item the coherence of \dict satisfies $\coh=\order(1/\sqrt{\outputdim})$
	\item the coherence of  \dicta satisfies $\coha=\order((\log\inputdim)/\outputdim)$
	\item the cardinality of \dicta satisfies $\inputdimA=\orderlower(\outputdim/(\log\inputdim))$.
\end{enumerate}

Similarly, we get from~\eqref{eq:condB} that\footnote{Note that the obvious condition $\nb\leq\inputdimB$ is implied by $\nb\leq \inputdimB/\spectralnorm{\dictb}^2$ since $\spectralnorm{\dictb}\geq1$.}
\be\label{eq:upperB}
	\elemB\leq c\min\lefto\{\frac{\cohb^{-2}}{\log\inputdim},\frac{\inputdimB}{\spectralnorm{\dictb}^2},\frac{\inputdimB}{\spectralnorm{\dicta}^2\spectralnorm{\dictb}^2}\right\}.
\ee
This upper bound is compatible with the scaling behavior $\elemB=\scales(\outputdim/(\log\inputdim))$ if and only if all of the following conditions are met:
\begin{enumerate}
\setcounter{enumi}{3}
	\item the coherence of \dictb satisfies $\cohb=\order(1/\sqrt{\outputdim})$
	\item the spectral norm of \dictb satisfies $\spectralnorm{\dictb}^2\leq c\,\inputdimB(\log\inputdim)/\outputdim$
	\item the spectral norm of \dicta satisfies $\spectralnorm{\dicta}^2\leq c\,\inputdimB(\log\inputdim)/(\spectralnorm{\dictb}^2\!\outputdim)$.
\end{enumerate}
%
%
Note that iv) is implied by i) since $\cohb\leq\coh$, by assumption.
Finally, it follows from i) that conditions~\eqref{eq:cond_l0} and~\eqref{eq:cond_l1} are compatible with the scaling behavior $\elemA=\scales(\outputdim/(\log\inputdim))$ and $\elemB=\scales(\outputdim/(\log\inputdim))$.

In the special case of \dicta and \dictb being  ONBs for $\complexset^\outputdim$, conditions ii) -- vi) are trivially satisfied. Hence, in the two-ONB case the square-root bottleneck is broken by randomizing according to the specifications in~\fref{thm:Tropp_ONB} whenever $\coh=\order(1/\sqrt{\outputdim})$, as already shown in~\cite{tropp2008}. The additional requirements ii) -- vi) become relevant for general dictionaries~\dict only.

We next present an example of a non-trivial dictionary \dict with sub-dictionaries \dicta and \dictb (not both ONBs)  that satisfy i) -- vi).  Let $\outputdim=p^k$, with $p$ prime and $k\in\naturals^+$. For this choice of \outputdim it is possible to design $\outputdim+1$ ONBs for $\complexset^\outputdim$, which, upon concatenation, form a dictionary \dict with coherence \coh equal to $1/\sqrt{\outputdim}$~\cite{calderbank1997,strohmer2003,gribonval2003}. In particular, the absolute value of the inner product between two distinct columns of \dict is, by construction, either $0$ or $1/\sqrt{\outputdim}$.
Obviously, for such a dictionary i) is satisfied. Furthermore, identifying \dicta with one of the $\outputdim+1$ ONBs and \dictb with the concatenation of the remaining \outputdim ONBs, we have $\coha=0$ and $\inputdimA=\outputdim$. Hence ii) and iii) are satisfied. 
Since \dictb consists of the concatenation of the remaining \outputdim ONBs, it has coherence $\cohb=1/\sqrt{\outputdim}$, and, hence, iv) is satisfied. Moreover, since \dictb is the concatenation of \outputdim ONBs for $\complexset^\outputdim$, it forms a tight frame for $\complexset^\outputdim$.
For a tight frame \dictb with  $\inputdimB=\outputdim^2$  frame elements in $\complexset^\outputdim$ (all \elltwo-normalized to one) the nonzero eigenvalues of the Gram matrix $\dictb^H\dictb$ are all equal to $\inputdimB/\outputdim=\outputdim$. Hence, the spectral norm of \dictb satisfies $\spectralnorm{\dictb}^2=\outputdim$.
Thus, v) is met. 
Finally, since \dicta is an ONB, its spectral norm satisfies $\spectralnorm{\dicta}^2=1$ and, therefore, condition vi) is met. 
%
%
%
Now, as a consequence of~\fref{thm:l0_l1_unique}, we obtain a robust sparsity threshold $\sparsity=\na+\nb$ of order $\outputdim/(\log\inputdim)$ for the dictionary $\dict=[\dicta\,\,\dictb]$. This threshold does not require that the positions of all nonzero entries of \inputvec are chosen randomly.
Specifically,  it suffices to randomize over the positions of the nonzero entries of \inputvec corresponding to \dictb, while the positions of the nonzero entries corresponding to \dicta can be chosen arbitrarily. 
%
%
As for the two-ONB case, once the support set of \inputvec is chosen, the values of all nonzero entries of \inputvec need to be chosen at random. 
%
%

Finally, as every dictionary $\dict\in\dictsetgenp$ can be viewed as the concatenation of two general dictionaries \dicta and \dictb such that $\dict=[\dicta\,\,\dictb]$, we can now ask the following question: Given a general dictionary \dict, over which part of the dictionary do we need to randomize to ``weed out'' the sparsity patterns that prohibit breaking the square-root bottleneck?
From the results above we obtain the intuitive answer that in the ``low-coherence'' part of the dictionary, namely \dicta,  we can pick the sparsity pattern arbitrarily, whereas the ``high-coherence'' part of the dictionary, namely \dictb, requires randomization. Note that, due to the bounds on the coherence parameters \coha and \cohb in ii) and iv), respectively, the ``low-coherence'' part \dicta of the overall dictionary \dict has, in general, fewer elements than the ``high-coherence'' part \dictb.  
%
Conditions i) -- vi) can be used to identify the largest possible part \dicta of the overall dictionary \dict where the corresponding sparsity pattern can be picked arbitrarily. Note, however, that the task of identifying the largest possible part \dicta is in general difficult.
\section{Conclusion}\label{sec:sum}
%
We presented a generalization of the uncertainty relation for the representation of a signal in two different ONBs~\cite{elad2002} to the representation of a signal in two general (possibly redundant or incomplete) signal sets.
This novel uncertainty relation is important in the context of the analysis of signals containing two distinct features each of which can be described sparsely only in an overcomplete signal set.
As shown in~\cite{studer2010}, the general uncertainty relation reported in this paper also forms  the basis for establishing recovery guarantees for signals that are sparse in a (possibly overcomplete) dictionary and corrupted by noise that is also sparse in a (possibly overcomplete) dictionary.

We furthermore presented a novel deterministic sparsity threshold guaranteeing uniqueness of the  (\Pzero)-solution for general dictionaries $\dict\in\dictsetp$, as well as thresholds guaranteeing equivalence of this unique (\Pzero)-solution to the solution obtained through BP and OMP. These thresholds improve on those previously known by up to a factor of two. Moreover, the known sparsity thresholds for general dictionaries and those for the concatenation of two ONBs follow from our results  as special cases.

Finally, the probabilistic recovery guarantees presented in this paper allow us to understand which parts of a general dictionary one needs to randomize over to ``weed out'' the sparsity patterns that prohibit breaking the square-root bottleneck.

\appendices

\section{Proof of~\fref{lem:uncertainty}}\label{app:uncertainty_lemma}
Assume that $\vecs\in\complexset^\outputdim$ can be represented as a linear combination of~$\na$ columns of~\dicta and, equivalently, as a linear combination of~$\nb$ columns of~\dictb. This means that there exists a vector \vecp with \na nonzero entries and a vector \vecq with \nb nonzero entries such that
\be\label{eq:uncertain1}
	\vecs=\dicta\vecp=\dictb\vecq.
\ee
We exclude the trivial case $\na=\nb=0$ and note that for $\na=0$ or $\nb=0$ we have $\vecs=\mathbf{0}_\outputdim$, by definition. 
%
%
%

Left-multiplication in~\eqref{eq:uncertain1} by $\dicta^H$ yields
\be\label{eq:uncertain2}
	\dicta^H\dicta\vecp=\dicta^H\dictb\vecq.
\ee
We next lower-bound the absolute value of the $i$th entry ($\allonotwo{i}{\inputdimA}$) of the vector $\dicta^H\dicta\vecp$ according to
\ba
	\abs{\left[\dicta^H\dicta\vecp\right]_i}&=\abs{[\vecp]_i+\sum_{j\neq i}\left[\dicta^H\dicta\right]_{i,j}[\vecp]_j}\nonumber\\
	&\geq \abs{[\vecp]_i}-\coha\sum_{j\neq i}\abs{[\vecp]_j}\label{eq:uncertain3a}\\
	&=(1+\coha)\abs{[\vecp]_i}-\coha\normone{\vecp}\label{eq:uncertain3}
\ea
where~\eqref{eq:uncertain3a} follows from the reverse triangle inequality and the fact that the off-diagonal entries of $\dicta^H\dicta$ can be upper-bounded in absolute value by \coha. Next, we upper-bound the absolute value of the $i$th entry of the vector $\dicta^H\dictb\vecq$ as follows:
\be\label{eq:uncertain4}
	\abs{\left[\dicta^H\dictb\vecq\right]_i}\leq\coh\normone{\vecq}.
\ee
Combining~\eqref{eq:uncertain3} and~\eqref{eq:uncertain4} yields
\be
	(1+\coha)\abs{[\vecp]_i}-\coha\normone{\vecp}\leq\coh\normone{\vecq}.
\een
If we now sum over all $i$ for which $[\vecp]_i\neq0$, we obtain
\be\label{eq:uncertain5pre}
	[(1+\coha)-\na\coha]\normone{\vecp}\leq\na\coh\normone{\vecq}
\ee
where we used that $\normzero{\vecp}=\na$, by assumption. Since $\na\coh\normone{\vecq}\geq0$, we can replace the LHS of~\eqref{eq:uncertain5pre} by the tighter bound
\be\label{eq:uncertain5}
	\pos{(1+\coha)-\na\coha}\normone{\vecp}\leq\na\coh\normone{\vecq}.
\ee
Multiplying both sides of~\eqref{eq:uncertain1} by $\dictb^H$ and following steps similar to the ones used to arrive at~\eqref{eq:uncertain5} yields
\be\label{eq:uncertain6}
	\pos{(1+\cohb)-\nb\cohb}\normone{\vecq}\leq\nb\coh\normone{\vecp}.
\ee
We now have to distinguish three cases.
If both $\na\geq1$ and $\nb\geq1$, and, hence, $\normone{\vecp}>0$ and $\normone{\vecq}>0$, we can combine~\eqref{eq:uncertain5} and~\eqref{eq:uncertain6} to obtain
\be\label{eq:uncertain_final}
	\na\nb\coh^2\geq\pos{(1+\coha)-\na\coha}\pos{(1+\cohb)-\nb\cohb}.
\ee
If $\na=0$ and $\nb\geq1$ (i.e., $\normone{\vecp}=0$ and $\normone{\vecq}>0$), we get from~\eqref{eq:uncertain6} that
\be\label{eq:uncertain7}
	\nb\geq1+\frac{1}{\cohb}.
\ee
Similarly, if $\nb=0$ and $\na\geq1$ (i.e., $\normone{\vecq}=0$ and $\normone{\vecp}>0$), we obtain from~\eqref{eq:uncertain5} that
\be\label{eq:uncertain8}
	\na\geq1+\frac{1}{\coha}.
\ee
Both~\eqref{eq:uncertain7} and~\eqref{eq:uncertain8} are contained in~\eqref{eq:uncertain_final} as special cases, as is easily verified. 
%
\section{Proof of~\fref{thm:spark_bound}}\label{app:spark_bound}
The proof will be effected by deriving a lower bound on the spark of dictionaries in \dictsetp, which together with~\eqref{eq:sparkP0}, yields the desired result~\eqref{eq:P0bound}.
This will be accomplished by finding a lower bound on the minimum number of nonzero entries that a nonzero vector $\vecv\in\complexset^\inputdim$ in the kernel of $\dict=[\dicta\,\,\dictb]$ must have. 
Without loss of generality, we may view \vecv as the concatenation of two vectors $\vecp\in\complexset^\inputdimA$ and $\vecq\in\complexset^\inputdimB$, i.e., $\vecv=[\vecp^T\,\,\vecq^T]^T$. As \vecv is in the kernel of $\dict=[\dicta\,\,\dictb]$, we have
\be
	[\dicta\,\,\dictb]\left[\begin{array}{c} \vecp\\ \vecq\end{array}\right] = \mathbf{0}_\outputdim.
\een
Therefore, the vectors \vecp and \vecq satisfy $\dicta\vecp=\dictb(-\vecq)\triangleq\vecs$. 
Let $\na\triangleq\normzero{\vecp}$ and $\nb\triangleq\normzero{-\vecq}=\normzero{\vecq}$ and 
recall that $\na=0$ is equivalent to $\vecp=\mathbf{0}_\inputdimA$ and $\nb=0$ is equivalent to $\vecq=\mathbf{0}_\inputdimB$, both by definition.
Since we require \vecv to be a nonzero vector, the case of $\na=\nb=0$ (and hence $\vecp=\mathbf{0}_\inputdimA$ and $\vecq=\mathbf{0}_\inputdimB$, and, therefore $\vecv=\mathbf{0}_\inputdim$) is excluded. 
For all other cases, the uncertainty relation in~\fref{lem:uncertainty} requires that the number of nonzero entries in \vecp and $-\vecq$ (representing \vecs according to $\dicta\vecp=\dictb(-\vecq)=\vecs$) satisfy 
%
%
%
\be\label{eq:uncertainty_app}
	\na\nb  \geq \dfrac{\pos{1-\coha(\na-1)}\pos{1-\cohb(\nb-1)}}{\coh^2}.
\ee
Based on~\eqref{eq:uncertainty_app}, we now derive a lower bound on $\spark(\dict)$ by considering the following three different cases: 
\paragraph*{The case $\nb\geq1$ and $\na=0$}
%
%
In this case, the vector $\vecv=[\vecp^T\,\,\vecq^T]^T$ in the kernel of $\dict=[\dicta\,\,\dictb]$ has nonzero entries only in the part $\vecq$ corresponding to sub-dictionary \dictb. 
%
It follows directly from~\eqref{eq:uncertainty_app} that
%
\be\label{eq:sparkb}
	\nb\geq1+\frac{1}{\cohb}.
\ee
\paragraph*{The case $\na\geq1$ and $\nb=0$}
%
%
In this case, the vector $\vecv=[\vecp^T\,\,\vecq^T]^T$ in the kernel of $\dict=[\dicta\,\,\dictb]$ has nonzero entries only in the part $\vecp$ corresponding to sub-dictionary \dicta. 
%
Again, direct application of~\eqref{eq:uncertainty_app} yields
\be\label{eq:sparka}
	\na\geq1+\frac{1}{\coha}.
\ee
%
\paragraph*{The case $\na\geq1$ and $\nb\geq1$}
In this case,  the vector $\vecv=[\vecp^T\,\,\vecq^T]^T$ in the kernel of $\dict=[\dicta\,\,\dictb]$ has nonzero entries in both parts \vecp and \vecq corresponding to sub-dictionary \dicta and \dictb, respectively.
%
Let \licardold denote the smallest possible number of nonzero entries of \vecv in this case. 
%
%
%
Together with~\eqref{eq:sparkb} and~\eqref{eq:sparka} we now have
\ba
	\spark(\dict)&\geq \min\left\{1+\frac{1}{\cohb},1+\frac{1}{\coha},\licardold\right\}\notag\\
	&= \min\left\{1+\frac{1}{\cohb},\licardold\right\}\label{eq:lb_spark}
\ea
where we used that $\coha\leq\cohb$, by assumption.
We next derive a lower bound on \licardold that is explicit in \coh, \coha, and \cohb. 
Specifically, we minimize $\na+\nb$ over all pairs $(\na,\nb)$ (with $\na\geq1$ and $\nb\geq1$) that satisfy~\eqref{eq:uncertainty_app}.
Since, eventually, we are interested in finding a lower bound on $\spark(\dict)$, it follows from~\eqref{eq:lb_spark} that it suffices to restrict the minimization to those pairs $(\na,\nb)$, for which both $\na\leq1+1/\cohb$ and $\nb\leq1+1/\cohb$. This implies that $[1-\coha(\na-1)]\geq0$ and $[1-\cohb(\nb-1)]\geq0$, and we thus have from~\eqref{eq:uncertainty_app} that
%
%
%
%
\begin{align}
	%
	\na\nb&\geq  \dfrac{[1-\coha(\na-1)][1-\cohb(\nb-1)]}{\coh^2}. \label{eq:uncertainty_app2}
\end{align}
%
%
Solving~\eqref{eq:uncertainty_app2} for \na, we get
\be
 \na \geq \dfrac{(1+\coha)(1+\cohb)-\nb\cohb(1+\coha)}{\nb(\coh^2-\coha\cohb)+\coha(1+\cohb)}\triangleq f(\nb).
\een
Finally, adding~\nb on both sides yields
\be
\label{eq:first_lb_on_licard}
	\na+\nb\geq f(\nb)+\nb. 
\ee
To arrive at a lower bound on $\na+\nb$ that is explicit in \coh, \coha, and \cohb only (in particular, the lower bound should be independent of \na and \nb), we further lower-bound the RHS of~\eqref{eq:first_lb_on_licard} by minimizing $f(\nb)+\nb$ as a function of $\nb$, under the constraints   $\na\geq 1$ and $\nb\geq 1$ (implied by assumption). 
This yields the following lower bound on \licardold:\footnote{The constraints $\na\geq f(\nb)$ and $\na\geq1$ are combined into  $\na\geq\max\{f(\nb),1\}$.}
\be
	\licardold\geq \min_{\nb\geq 1} [\max\{f(\nb),1\}+\nb]\triangleq\licard.
\een
We now have that
\be\label{eq:minimumislower}
\begin{split}
\licard & = \min_{\nb\geq 1} [\max\{f(\nb),1\}+\nb]\\
&\leq [\max\{f(\nb),1\}+\nb]\bigr|_{\nb=1/\cohb}\\
&= 1+\frac{1}{\cohb}
\end{split}
\ee
where we used the fact that $f(1/\cohb)\leq 1$.
%
%
%
As a consequence of~\eqref{eq:lb_spark}, the inequality in~\eqref{eq:minimumislower} implies that 
\ba
\label{eq:spark_final}
\spark(\dict)&\geq\licard\notag\\
&= \min_{\nb\geq 1} [\max\{f(\nb),1\}+\nb]\nonumber\\
&\geq \min_{x\geq 1} \,[\max\{f(x),1\}+x]
\ea
where~\eqref{eq:spark_final} follows because minimizing over all $x\in\reals$ with $x\geq1$ yields a lower bound on the minimum taken over the integer parameter \nb only.
We next compute the minimum in~\eqref{eq:spark_final}. 
The function $f(x)$ can be shown to be strictly decreasing. Furthermore, the equation $f(x)=1$ has the unique solution~$\xbord=(1+\cohb)\big/(\cohb+\coh^2)\geq 1$,
%
%
where the inequality follows because $\coh\leq 1$, by definition. 
We can therefore rewrite~\eqref{eq:spark_final} as
\be
\label{eq:final_min}
\min_{x\geq 1} [\max\{f(x),1\}+x]
=\min_{1\leq x \leq \xbord} [f(x)+x].
\ee
In the case $\coha=\cohb=\coh$, the function $g(x)\triangleq f(x)+x$ reduces to the constant $1+1/\coh$ so that $\spark(\dict)\geq1+1/\coh$.
In all other cases, 
the function $g(x)$ is strictly convex for $x\geq 0$. 
Furthermore, we have~$g(1)\geq g(\xbord)$ as a consequence of the assumption~$\coha\leq \cohb$.   
Hence, the minimum in~\eqref{eq:final_min} is attained either at the boundary point~\xbord, or at the stationary point \xstat of $g(x)$,
which is given by
\be\label{eq:xstat}
	\xstat=\frac{\coh\sqrt{(1+\coha)(1+\cohb)}-\coha(1+\cohb)}{\coh^2-\coha\cohb}\geq 1.
\ee
The inequality in~\eqref{eq:xstat} follows from the convexity of $g(x)$ and the fact that $g(1)\geq g(\xbord)$.
If the stationary point \xstat is inside the interval $[1,\xbord]$, the minimum is attained at $\hat{x}=\xstat$, otherwise it is attained at $\hat{x}=\xbord$.

\section{The Sparsity Threshold in~\fref{thm:spark_bound} Improves on the Threshold in~\eqref{eq:gen_sparsity_th}}\label{app:betterSparkbound}
%
We show that the threshold in~\eqref{eq:P0bound} improves on that in~\eqref{eq:gen_sparsity_th}, unless $\cohb=\coh$ or $\coh=1$, in which case the threshold in~\eqref{eq:P0bound} is the same as that in~\eqref{eq:gen_sparsity_th}. 
This will be accomplished by considering the two (mutually exclusive) cases $\xbord\leq\xstat$ and $\xbord>\xstat$.
\paragraph*{The case $\xbord\leq \xstat$} The threshold in~\eqref{eq:P0bound} equals
\be
	\frac{f(\xsol)+\xsol}{2}=\frac{f(\xbord)+\xbord}{2}= \frac{1}{2}\lefto(1 + \dfrac{1+\cohb}{\cohb+\coh^2}\right).
\een
It is now easily verified that 
\be
	\frac{1}{2}\lefto(1 + \dfrac{1+\cohb}{\cohb+\coh^2}\right)\geq\frac{1}{2}\lefto(1+\frac{1}{\coh}\right)
\een
for all $\cohb\leq\coh\leq 1$ with equality if and only if $\cohb=\coh$ or $\coh=1$. Note that for $\cohb=\coh$ (irrespective of \coha) or for $\coh=1$ (irrespective of \coha and \cohb),  we have $\xbord\leq\xstat$. 
%
\paragraph*{The case $\xbord >\xstat$} Set~$\Delta=\sqrt{(1+\coha)(1+\cohb)}-\coh$.
The function~$f(\xstat)+\xstat$, which we denote as $h(\coha,\cohb,\coh)$ to highlight its dependency on the variables~\coha, \cohb, and~\coh, is strictly decreasing in~\coha (for fixed \cohb and \coh) as long as 
$\cohb/\coh<\Delta<\coh/\cohb$.
Since $\xbord>\xstat$ implies $\cohb<\coh$, and since $\coha\leq\cohb$, by assumption, the inequality $\Delta<\coh/\cohb$ is always satisfied.
The inequality~$\cohb/\coh<\Delta$ holds whenever $\xstat<\xbord$, which is satisfied by assumption. 
Hence, we have that 
\ba
	\frac{f(\xstat)+\xstat}{2}=\frac{h(\coha,\cohb,\coh)}{2} &\geq \frac{h(\cohb,\cohb,\coh)}{2}\label{eq:equality1} \\
	&=\dfrac{1+\cohb}{\coh+\cohb}\notag\\
	&\geq\frac{1}{2}\lefto(1+\frac{1}{\coh}\right).\label{eq:equality2}
\ea
Note that equality in~\eqref{eq:equality1} and~\eqref{eq:equality2} holds if and only if~$\coha=\cohb=\coh$, already treated in the case $\xbord\leq\xstat$.

\section{Proof of~\fref{thm:l1_na_nb}}\label{app:l1proof}
%
Our proof essentially follows the program laid out in~\cite{tropp2003} for dictionaries in \dictsetonbp, with appropriate modifications to account for the fact that we are dealing with the concatenation of two general dictionaries.
Let~\subD be the matrix that contains the columns of \dicta and \dictb participating in the representation of $\outputvec=[\dicta\,\,\dictb]\,\inputvec$, i.e., the columns in $[\dicta\,\,\dictb]$ corresponding to the nonzero entries in \inputvec.
A sufficient condition for BP \emph{and} OMP applied to $\outputvec=[\dicta\,\,\dictb]\,\inputvec$ to recover \inputvec is~\cite[Thm. 3.1, Thm. 3.3]{tropp2004}
\be\label{eq:ERC}
	\max_{\cold_i}\normone{\subD^\dagger\cold_i}<1
\ee
where the maximization in~\eqref{eq:ERC} is performed over all columns $\cold_i$ in \dict that do not appear in \subD. 
%
We prove the theorem by first carefully bounding the absolute value of each element of the vector $\subD^\dagger\cold_i$.
Concretely, we start with the following inequality
\begin{align*}
	\abs{[\subD^\dagger\cold_i]_k}&=\abs{[(\herm{\subD}\subD)^{-1} \herm{\subD}\cold_i]_k}\\
	&= \abs{\sum_l [(\herm{\subD}\subD)^{-1}]_{k,l} [\herm{\subD}\cold_i]_l} \\
	&\leq \sum_l \abs{[(\herm{\subD}\subD)^{-1}]_{k,l}} \abs {[\herm{\subD}\cold_i]_l}
\end{align*}
and then bound the absolute value of each entry of the matrix $(\herm{\subD}\subD)^{-1}$ and of each element of the vector $\herm{\subD}\cold_i$. We will verify below that the matrix $\herm{\subD}\subD$ is invertible.
To simplify notation, for any matrix \matA, we let $\abs{\matA}$ be the matrix with entries
\begin{align*}
	[\abs{\matA}]_{k,l} = \abs{[\matA]_{k,l}}.
\end{align*}
Furthermore, if for two matrices \matA and \matB of the same size we have that
\begin{align*}
	\abs{[\matA]_{k,l}} \leq \abs{[\matB]_{k,l}}
\end{align*}
for all  pairs $(k,l)$, we shall write $\abs{\matA} \leqentry \abs{\matB}$.

\subsection{Bound on the Elements of $(\herm{\subD}\subD)^{-1}$} 
\label{sec:entry_wise_bound_on_}
Since the columns of \dict are \elltwo-normalized to $1$, we can write
\be
	\subD^H\subD = \bI_{\elemA+\elemB}-\bK
\een 
where $-\bK$ contains the off-diagonal elements of $\subD^H\subD$.
Clearly,
%
\begin{align}
	\label{eq:bound_on_K}
	\abs{\bK} &\leqentry \mat 
			\coha(\ones_{\elemA,\elemA}-\bI_{\elemA})& \coh\ones_{\elemA,\elemB} \\
			\coh \ones_{\elemB,\elemA} &	\cohb(\ones_{\elemB,\elemB}-\bI_{\elemB})
	 					\emat
	\leqentry
			\mat 
			\cohb(\ones_{\elemA,\elemA}-\bI_{\elemA})& \coh\ones_{\elemA,\elemB} \\
			\coh \ones_{\elemB,\elemA} &	\cohb(\ones_{\elemB,\elemB}-\bI_{\elemB})
	 		\emat \notag\\
	&= -\cohb \bI_{\elemA+\elemB} + \coh \ones_{\elemA+\elemB, \elemA+\elemB} - (\coh -\cohb)\bT
\end{align}
where we set
\begin{align*}
	\bT = \mat
	\ones_{\elemA,\elemA} & \zeroes_{\elemA,\elemB}\\
	\zeroes_{\elemB,\elemA}&\ones_{\elemB,\elemB}
	\emat.
\end{align*}
As a consequence of~\eqref{eq:bound_on_K} and using the assumption $\elemB\geq\elemA$, we have that $\vecnorm{\bK}_{1,1}\leq \coh\elemB + \cohb(\elemA-1)$.
Since $\vecnorm{\cdot}_{1,1}$ is a matrix norm~\cite[p. 294]{hornjohnson}, the matrix $\subD^H\subD$ is invertible whenever $\coh\elemB + \cohb(\elemA-1)<1$, and, moreover, we can expand $(\subD^H\subD)^{-1}$ into a Neumann series  according to $(\subD^H\subD)^{-1}=\bI_{\elemA+\elemB}+\sum_{k=1}^\infty \bK^k$.
%
As the condition in~\eqref{eq:BP_OMP_recovery_na_nb} implies that  $\coh\elemB+\cohb(\elemA-1)<1$, we have
\begin{align}
	\abs{(\subD^H\subD)^{-1}}
	 &= \abs{\bI_{\elemA+\elemB}+\sum_{k=1}^\infty \bK^k}\label{eq:Neumann}\\
	&\leqentry \bI_{\elemA+\elemB} +\sum_{k=1}^\infty \abs{\bK}^k\label{eq:triangle_for_matrix}\\
	&\leqentry \bI_{\elemA+\elemB} +\sum_{k=1}^\infty [-\cohb \bI_{\elemA+\elemB} + \coh \ones_{\elemA+\elemB, \elemA+\elemB} - (\coh -\cohb)\bT]^k \notag\\
	&= \bigl[\underbrace{(1+\cohb)\bI_{\elemA+\elemB} + (\coh -\cohb)\bT}_{\triangleq\matX}-\coh\ones_{\elemA+\elemB,\elemA+\elemB}\bigr]^{-1} \notag\\
	&=\bigl[\matI_{\elemA+\elemB}- \coh\matX^{-1}\ones_{\elemA+\elemB,\elemA+\elemB}\bigr]^{-1}\matX^{-1}.
	\label{eq:upper-bound_matrix_inverse}
\end{align}
Here, in~\eqref{eq:triangle_for_matrix} we used the triangle inequality and the fact that $\abs{\bK^k}\leqentry\abs{\bK}^k$.
%
%
We next compute the inverses in~\eqref{eq:upper-bound_matrix_inverse}.
To get $\matX^{-1}$, we use the fact that \matX is a block-diagonal matrix and apply Woodbury's identity~\cite[p.19]{hornjohnson} to each of the two blocks,\footnote{To apply Woodbury's identity, we exploit the fact that $\mathbf{1}_{n,n}=\mathbf{1}_n\mathbf{1}_n^T$.} which yields
%
%
%
\begin{align}
	\label{eq:X_inverse_app}
	\matX^{-1}=
	\mat
		\dfrac{1}{1+\cohb}\left( \matI_{\elemA} - \dfrac{\coh-\cohb}{(\coh-\cohb)\elemA+1+\cohb} \ones_{\elemA,\elemA}\right) & \zeroes_{\elemA,\elemB} \\
	\zeroes_{\elemB,\elemA} & 	\dfrac{1}{1+\cohb}\left( \matI_{\elemB} - \dfrac{\coh-\cohb}{(\coh-\cohb)\elemB+1+\cohb}\ones_{\elemB,\elemB} \right)
	\emat.
\end{align}
Next, setting $\consta=[(\coh-\cohb)\elemA+1+\cohb]^{-1}$, $\constb=[(\coh-\cohb)\elemB+1+\cohb]^{-1}$, and
\begin{align*}
	\vecv=
	\coh
	\mat 
	\consta \ones_{\elemA} \\
	\constb \ones_{\elemB}
	 \emat
\end{align*}
steps similar to the ones reported in \cite[Eq. (A.2)-(A.3)]{tropp2003} yield
\begin{align}
	\label{eq:inverse_first_matrix}
	\bigl[\matI_{\elemA+\elemB}- \coh\matX^{-1}\ones_{\elemA+\elemB,\elemA+\elemB}\bigr]^{-1} =\matI_{\elemA+\elemB} +\frac{1}{1-\coh(\consta\elemA+\constb\elemB)} \vecv\tp{\ones_{\elemA+\elemB}}.
\end{align}
Using the fact, shown in~\eqref{eq:upper-bound_matrix_inverse}, that 
\be
	\abs{(\subD^H\subD)^{-1}}\leqentry\bigl[\matI_{\elemA+\elemB}- \coh\matX^{-1}\ones_{\elemA+\elemB,\elemA+\elemB}\bigr]^{-1}\matX^{-1}
\een
we can combine~\eqref{eq:X_inverse_app} and~\eqref{eq:inverse_first_matrix} to obtain an upper bound on the absolute value of each entry of $(\herm{\subD}\subD)^{-1}$.
\subsection{Bound on the Elements of $\herm{\subD}\cold_i$} 
\label{sec:entry_wise_bound_on_vector}
Let $\cold_i$ be a column of \dict that does not appear in \subD.
Assume that $\cold_i \in \dicta$ (we will later show that in searching the maximum in~\eqref{eq:ERC} it is, indeed, sufficient to assume  $\cold_i\in\dicta$).
Then, we have
\begin{align}\label{eq:vector_bound}
	\abs{\herm{\subD}\cold_i}\leqentry 
	\mat
	\coha \ones_{\elemA} \\
	\coh  \ones_{\elemB} \emat \leqentry 
	\mat
	\cohb \ones_{\elemA} \\
	\coh  \ones_{\elemB} 
	\emat.
\end{align}
As a sideremark, we note that  we loose the dependency of our final result on~\coha through the bounds~\eqref{eq:bound_on_K} and~\eqref{eq:vector_bound}. 
%
\subsection{Putting the Pieces Together} 
\label{sec:putting_pieces_together}
Substituting~\eqref{eq:inverse_first_matrix} into~\eqref{eq:upper-bound_matrix_inverse}, we get
\begin{align}
	\abs{\subD^\dagger\cold_i} &\leqentry 
	\left(\matI_{\elemA+\elemB} +\frac{1}{1-\coh(\consta\elemA+\constb\elemB)} \vecv\tp{\ones_{\elemA+\elemB}}\right)
	\matX^{-1} \mat
	\cohb \ones_{\elemA} \\
	\coh  \ones_{\elemB} 
	\emat \notag\\
	&=\left(\matI_{\elemA+\elemB} +\frac{1}{1-\coh(\consta\elemA+\constb\elemB)} \vecv\tp{\ones_{\elemA+\elemB}}\right) 
	\mat
	\cohb\consta \ones_{\elemA} \\
	\coh\constb  \ones_{\elemB}
	\emat\notag\\
	&= 
	\frac{1}{1-\coh(\consta\elemA+\constb\elemB)}
	\mat
	(\cohb\consta+(\coh-\cohb)\coh\nb\consta\constb) \ones_{\elemA} \\
	(\coh\constb-(\coh-\cohb)\coh\na\consta\constb) \ones_{\elemB}
	\emat.\label{eq:vector_abs_value} 
\end{align}
%
%
Summing the RHS of~\eqref{eq:vector_abs_value} over all entries of the vector $\subD^\dagger\cold_i$ yields the following upper bound on~$\normone{\subD^\dagger\cold_i}$:
\begin{align}\label{eq:result_app_d}
	\normone{\subD^\dagger\cold_i} \leq \frac{\cohb\consta\elemA+\coh\constb\elemB}{1-\coh(\consta\elemA+\constb\elemB)}.
\end{align}

If we instead assume that $\cold_i \in \dictb$ and apply the same steps as before, we find that 
\begin{align}
	\label{eq:alt_result_app_d}
	\normone{\subD^\dagger\cold_i} \leq\frac{\coh\consta\elemA+\cohb\constb\elemB}{1-\coh(\consta\elemA+\constb\elemB)}.
\end{align}
%
%
%
Since $\cohb\consta\elemA+\coh\constb\elemB\geq \coh\consta\elemA+\cohb\constb\elemB$ it follows that 
\be
	 \frac{\coh\consta\elemA+\cohb\constb\elemB}{1-\coh(\consta\elemA+\constb\elemB)}\leq\frac{\cohb\consta\elemA+\coh\constb\elemB}{1-\coh(\consta\elemA+\constb\elemB)}
\een
and hence
\ba
	\max_{\cold_i}\normone{\subD^\dagger\cold_i}
	&\leq\frac{\cohb\consta\elemA+\coh\constb\elemB}{1-\coh(\consta\elemA+\constb\elemB)}.\notag
\ea	
We can therefore conclude that a sufficient condition for BP and OMP applied to $\outputvec=\dict\inputvec$ to recover~\inputvec~is
\begin{align}
	\label{eq:final_result_app_d}
	\frac{\cohb\consta\elemA+\coh\constb\elemB}{1-\coh(\consta\elemA+\constb\elemB)}< 1.
\end{align}
Simple algebraic manipulations reveal that~\eqref{eq:final_result_app_d} is equivalent to~\eqref{eq:BP_OMP_recovery_na_nb}.
%

%
%
%

%
%
%

%
%
%
%

\section{Proof of \fref{cor:l1_general}}\label{app:proof_cor_BP_OMP} 
\label{sec:proof_of_cor_l1_general}
%
%
We obtain~\fref{cor:l1_general} as a consequence of~\fref{thm:l1_na_nb} as follows.
For given $\nb\geq\na$ it follows from~\eqref{eq:BP_OMP_recovery_na_nb} that a sufficient condition for BP and OMP to recover the unknown vector \inputvec is
\be
	\na<\frac{(1+\cohb)^2-\nb(1+\cohb)(\coh+\cohb)}{2\cohb(1+\cohb)+2\nb(\coh^2-\cohb^2)}\triangleq h(\nb).
\een
%
%
To arrive at a sparsity threshold that is explicit in \cohb and \coh only, we minimize $h(\nb)+\nb$ over \nb, under the constraint $\elemB\geq 1$ (recall that $\nb\geq\na$ and note that representing a nonzero vector $\outputvec\in\complexset^\outputdim$ requires at least one column of \dict). 
Furthermore, we have that 
%
%
\begin{align}
	\label{eq:maximization_app_e}
	\min_{\nb\geq 1}[h(\nb)+\nb]\geq\min_{x\geq 1}[h(x)+x]\triangleq \sparsity
\end{align}
where $x\in\reals$. Clearly, minimizing over all $x\geq1$ with $x\in\reals$, as opposed to integer values \nb only, can only yield a smaller value for the minimum.  In the case $\cohb=\coh$, the function $h(x)+x$ reduces to the constant $(1+1/\coh)/2$, thereby recovering the previously known sparsity threshold in~\eqref{eq:gen_sparsity_th}. 
In all other cases, the function $h(x)+x$ is strictly convex for $x\geq0$. Hence, the minimum in~\eqref{eq:maximization_app_e} is attained either at the boundary point $x=1$ or at the stationary point $x_s$ of $h(x)+x$, given by
\be
	x_s=\frac{(1+\cohb)(\sqrt{2\coh(\cohb+\coh)}-2\cohb)}{2(\coh^2-\cohb^2)}.
\een
If the stationary point satisfies $x_s>1$, then the minimum in~\eqref{eq:maximization_app_e} is attained at the stationary point, otherwise the minimum is attained at the boundary point $x=1$.
The condition $x_s>1$ is equivalent to the condition $\thlone>1$ (where $\thlone$ is defined in~\eqref{eq:condition_corollary_omp_bp}). If $\thlone>1$ the minimum in~\eqref{eq:maximization_app_e} is given by
\be
	\sparsity= \dfrac{(1+\cohb)\bigl[2\sqrt{2}\sqrt{\coh(\cohb+\coh)}-(\coh+3\cohb)\bigr]}{2(\coh^2-\cohb^2)}.
\een
If $\thlone\leq1$, the minimum in~\eqref{eq:maximization_app_e} is attained at the boundary point $x=1$ and is given by
\begin{align}\label{eq:sparsityTH2}
	\sparsity= \dfrac{1+2\coh^2 +3\cohb -\coh(1+\cohb)}{2(\coh^2+\cohb)}.
\end{align}
Note that for $\cohb=\coh$ the sparsity threshold in~\eqref{eq:sparsityTH2} reduces to that in~\eqref{eq:gen_sparsity_th}.

\section{The Sparsity Threshold in~\fref{cor:l1_general} Improves on the Threshold in~\eqref{eq:gen_sparsity_th}}\label{app:proofl1better}
%
We show that the threshold in~\fref{cor:l1_general} improves on that in~\eqref{eq:gen_sparsity_th}, unless $\cohb=\coh$ or $\coh=1$, in which case the threshold in ~\fref{cor:l1_general}  is the same as that in~\eqref{eq:gen_sparsity_th}.  
Let us first consider the case when the RHS of~\eqref{eq:l1_final}  in~\fref{cor:l1_general} reduces to 
\be
	\sparsity \triangleq \frac{1+2\coh^2 +3\cohb -\coh(1+\cohb)}{2(\coh^2+\cohb)}.
\een
We need to establish that
\be\label{eq:l1better_part2}
	\frac{1+2\coh^2 +3\cohb -\coh(1+\cohb)}{2(\coh^2+\cohb)}\geq\frac{1}{2}\lefto(1+\frac{1}{\coh}\right)
\ee
with equality if and only if $\cohb=\coh$ or $\coh=1$.
Straightforward calculations reveal that the inequality~\eqref{eq:l1better_part2} is equivalent to
\be\label{eq:simple_cond}
	(\coh-\cohb)(1-\coh)^2\geq0
\ee
which is satisfied for all $\cohb\leq\coh$. Furthermore, equality in~\eqref{eq:simple_cond} holds if and only if $\cohb=\coh$ or $\coh=1$.

Next, we consider the case $\cohb<\coh$ and $\kappa(\coh,\cohb)>1$ so that the RHS of~\eqref{eq:l1_final} reduces to
\be
	\sparsity=\frac{(1+\cohb)\bigl[2\sqrt{2}\sqrt{\coh(\cohb+\coh)}-(\coh+3\cohb)\bigr]}{2(\coh^2-\cohb^2)}.
\een
For  $\coh\leq7/9$ it can be verified that
\be
	\frac{(1+\cohb)\bigl[2\sqrt{2}\sqrt{\coh(\cohb+\coh)}-(\coh+3\cohb)\bigr]}{2(\coh^2-\cohb^2)}>\frac{1}{2}\lefto(1+\frac{1}{\coh}\right).
\een
It turns out that a necessary condition for $\kappa(\coh,\cohb)>1$ is $\coh<1/\sqrt{2}$. 
The proof is completed by noting that $1/\sqrt{2}<7/9$.

\section{Proof of \fref{lem:subdictionary}}\label{app:proofs}
%
%
Since the minimum singular value $\sigma_\text{min}(\subD)$ of the sub-dictionary \subD can be lower-bounded as $\sigma_\text{min}^2(\subD)\geq1-\spectralnorm{\subD^H\subD-\matI_{\elemA+\elemB}}$, we have 
%
%
\ba
	\Prob\lefto\{\sigma_\text{min}(\subD)\leq  \frac{1}{\sqrt{2}}\right\} & = \Prob\lefto\{\sigma_\text{min}^2(\subD)\leq \frac{1}{2}\right\}\notag \\
	& \leq \Prob\lefto\{1-\spectralnorm{\subD^H\subD-\matI_{\elemA+\elemB}}\leq \frac{1}{2}\right\}\notag\\
	 &= \Prob\lefto\{\spectralnorm{\subD^H\subD-\matI_{\elemA+\elemB}}\geq \frac{1}{2}\right\}.\label{eq:singularVSnorm}
\ea
Next, we study the tail behavior of the random variable $\hollowS=\spectralnorm{\subD^H\subD-\matI_{\elemA+\elemB}}$, which will then allow us to upper-bound $\Prob\lefto\{\spectralnorm{\subD^H\subD-\matI_{\elemA+\elemB}}\geq 1/2\right\}$. To this end the following lemma, which follows from Markov's inequality, will be useful.
\begin{lem}[\!{\cite[Prop. 10]{tropp2008}}]\label{lem:tailbound}
If~the~mo\-ments of a nonnegative random variable $R$ can be upper-bounded as $[\Exop(R^q)]^{1/q}\leq\alpha\sqrt{q}+\beta$ for all $q\geq Q\geq1$, where $\alpha,\beta>0$, then, 
\be
	\Prob\{R\geq e^{1/4}(\alpha u+\beta)\}\leq e^{-u^2/4}
\een
for all $u\geq\sqrt{Q}$.
\end{lem}
\vspace{3mm}
To be able to apply~\fref{lem:tailbound} to $\hollowS=\spectralnorm{\subD^H\subD-\matI_{\elemA+\elemB}}$, we first need  an upper bound on $[\Exop(\hollowS^q)]^{1/q}$ that is of the form $\alpha\sqrt{q}+\beta$. To derive this upper bound, we start by writing \subD as $\subD = [\subA\,\,\subB]$,
%
where \subA and \subB denote the matrices containing the columns chosen arbitrarily from \dicta and randomly from \dictb, respectively.
We then obtain
\be
	\subD^H\subD-\matI_{\elemA+\elemB} = \lefto[\!\!\begin{array}{cc}\subA^H\subA-\matI_\elemA\!\!\!\! & \subA^H\subB\\ \subB^H\subA\!\!\!\! & \subB^H\subB-\matI_\elemB\end{array}\!\!\right].
\een
Applying the triangle inequality for operator norms, we can now upper-bound  $\hollowS$ according to
\ba
	\hollowS & = \spectralnorm{\!\lefto[\!\!\begin{array}{cc}\subA^H\subA-\matI_\elemA\!\!\!\! & \subA^H\subB\\ \subB^H\subA\!\!\!\! & \subB^H\subB-\matI_\elemB\end{array}\!\!\right]\!}\nonumber\\
	& \leq \spectralnorm{\!\lefto[\!\!\begin{array}{cc}\subA^H\subA-\matI_\elemA\!\!\!\! & \mathbf{0}\\ \mathbf{0} \!\!\!\!& \subB^H\subB-\matI_\elemB\end{array}\!\!\right]\!}+ \spectralnorm{\!\lefto[\!\!\begin{array}{cc}\mathbf{0} \!\!\!\!& \subA^H\subB\\ \subB^H\subA \!\!\!\!& \mathbf{0}\end{array}\!\!\right]\!}\nonumber\\
	%
	&\leq\max\lefto\{\spectralnorm{\subA^H\subA-\matI_\elemA},\spectralnorm{\subB^H\subB-\matI_\elemB}\right\}+\spectralnorm{\subA^H\subB}\nonumber\\
	&\leq\spectralnorm{\subA^H\subA-\matI_\elemA}+\spectralnorm{\subB^H\subB-\matI_\elemB}+\spectralnorm{\subA^H\subB}\label{eq:spectralsum}
\ea
%
%
where the second inequality follows because the spectral norm of both a block-diagonal matrix and an anti-block-diagonal matrix is given by the largest among the spectral norms of the individual nonzero blocks.
Next, we define $\hollowA=\spectralnorm{\subA^H\subA-\matI_\elemA}$, $\hollowB=\spectralnorm{\subB^H\subB-\matI_\elemB}$, and $\cross=\spectralnorm{\subA^H\subB}$.
It then follows from~\eqref{eq:spectralsum} that for all $q\geq 1$
\ba
	\lefto[\Exop (\hollowS^q)\right]^{1/q}&\leq\lefto[\Exop\lefto(\lefto(\hollowA+\hollowB+\cross\right)^q\right)\right]^{1/q}\nonumber\\
	&\leq \lefto[\Exop(\hollowA^q)\right]^{1/q}+\lefto[\Exop(\hollowB^q)\right]^{1/q}+\lefto[\Exop\lefto(\cross^q\right)\right]^{1/q}\nonumber\\
	&=\hollowA+\lefto[\Exop(\hollowB^q)\right]^{1/q}+\lefto[\Exop\lefto(\cross^q\right)\right]^{1/q}\label{eq:threeterms}
\ea
%
where the second inequality is a consequence of the triangle inequality for the norm $[\Exop(\abs{\cdot}^q)]^{1/q}$  (recall that we assumed $q\geq1$ and hence $[\Exop(\abs{\cdot}^q)]^{1/q}$ is a norm), and in the last step we used the fact that \hollowA is a deterministic quantity. All expectations in~\eqref{eq:threeterms} are with respect to the random choice of columns from the sub-dictionary \dictb. 

We next upper-bound the three terms on the RHS of \eqref{eq:threeterms} individually. Applying Ger\v{s}gorin's disc theorem~\cite[Thm.~6.1.1]{hornjohnson} to the first term, we obtain
\be
	 \hollowA=\spectralnorm{\subA^H\subA-\matI_\elemA} \leq (\elemA-1)\coha.\label{eq:firstterm}
\ee
For the second term, we use~\cite[Eq. (6.1)]{tropp2008} to get 
\ba\label{eq:tropp6.1}
	\lefto[\Exop(\hollowB^q)\right]^{1/q}&=\lefto[\Exop\lefto(\spectralnorm{\subB^H\subB-\matI_\elemB}^q\right)\right]^{1/q}\nonumber\\
	&\leq\sqrt{144\cohb^2\elemB r_1}+\frac{2\elemB}{\inputdimB}\spectralnorm{\dictb}^2
\ea
where $r_1=\max\lefto\{1,\log\lefto(\elemB/2+1\right)\!,q/4\right\}$.
Assuming that $q\geq\max\{4\log(\elemB/2+1),4\}$ and, hence, $r_1=q/4$, we can simplify~\eqref{eq:tropp6.1} to
\be\label{eq:secondterm}
	\lefto[\Exop(\hollowB^q)\right]^{1/q}\leq6\sqrt{\cohb^2\elemB}\sqrt{q}+\frac{2\elemB}{\inputdimB}\spectralnorm{\dictb}^2.
\ee
To bound the third term, we use the upper bound in~\cite[Thm. 8]{tropp2008} on the spectral norm of a random compression combined with the fact that $\rank(\subA^H\subB)\leq\elemB$, which is a consequence of $\subA^H\subB$ being of dimension $\na\times\nb$. This yields
\ba
	\lefto[\Exop\lefto(\cross^q\right)\right]^{1/q}&=\lefto[\Exop\lefto(\spectralnorm{\subA^H\subB}^q\right)\right]^{1/q}\nonumber\\
	&\leq3\sqrt{r_2}\,\vecnorm{\subA^H\dictb}_{1,2}+\sqrt{\frac{\elemB}{\inputdimB}}\spectralnorm{\subA^H\dictb}\label{eq:lastterm1}
\ea
where $r_2=\max\lefto\{2,2\log\elemB,q/2\right\}$.
Assuming that $q\geq\max\{4\log\elemB,4\}$, we can further upper-bound the RHS of~\eqref{eq:lastterm1} to get
\ba
	\lefto[\Exop\lefto(\cross^q\right)\right]^{1/q}&\leq\frac{3}{\sqrt{2}}\sqrt{q}\,\vecnorm{\subA^H\dictb}_{1,2}+\sqrt{\frac{\elemB}{\inputdimB}}\spectralnorm{\subA^H\dictb}\nonumber\\
	&\leq \frac{3}{\sqrt{2}}\sqrt{\coh^2\elemA}\sqrt{q}+\sqrt{\frac{\elemB}{\inputdimB}}\spectralnorm{\subA^H\dictb}\label{eq:lasttermcoh}\\
	&\leq\frac{3}{\sqrt{2}}\sqrt{\coh^2\elemA}\sqrt{q}+\sqrt{\frac{\elemB}{\inputdimB}}\spectralnorm{\dicta}\spectralnorm{\dictb}\label{eq:lasttermspectralnorm}
\ea
where~\eqref{eq:lasttermcoh} follows from the fact that the magnitude of each entry of $\subA^H\dictb$ is upper-bounded by $\coh$ and, thus, $\vecnorm{\subA^H\dictb}_{1,2}\leq \sqrt{\coh^2\elemA}$. To arrive at~\eqref{eq:lasttermspectralnorm} we used $\spectralnorm{\subA^H\dictb}\leq\spectralnorm{\subA^H}\spectralnorm{\dictb}\leq\spectralnorm{\dicta}\spectralnorm{\dictb}$, which follows from the sub-multiplicativity of the spectral norm and the fact that the spectral norm of the submatrix \subA of \dicta cannot exceed that of \dicta~\cite[Thm. 4.3.3]{hornjohnson}. 
We can now combine the upper bounds~\eqref{eq:firstterm},~\eqref{eq:secondterm}, and~\eqref{eq:lasttermspectralnorm} to obtain  
\ba
	\lefto[\Exop\lefto(\hollowS^q\right)\right]^{1/q}&\leq(\elemA-1)\coha+6\sqrt{\cohb^2\elemB}\sqrt{q}+\frac{2\elemB}{\inputdimB}\spectralnorm{\dictb}^2+\nonumber\\
	&\dummyrel{=}\,\,+\frac{3}{\sqrt{2}}\sqrt{\coh^2\elemA}\sqrt{q}+\sqrt{\frac{\elemB}{\inputdimB}}\spectralnorm{\dicta}\spectralnorm{\dictb}\nonumber\\
	& = \underbrace{\lefto(6\sqrt{\cohb^2\elemB}+\frac{3}{\sqrt{2}}\sqrt{\coh^2\elemA}\right)}_{\alpha}\sqrt{q}\,+\nonumber\\
	&\dummyrel{=}\,\,+\underbrace{(\elemA-1)\coha+\frac{2\elemB}{\inputdimB}\spectralnorm{\dictb}^2+\sqrt{\frac{\elemB}{\inputdimB}}\spectralnorm{\dicta}\spectralnorm{\dictb}}_{\beta}\nonumber\\
	& = \alpha\sqrt{q}+\beta\nonumber
\ea
for all $q\geq Q_1=\max\{4\log(\elemB/2+1),4\log\elemB,4\}$.
Hence,~\fref{lem:tailbound} yields
\be
	\Prob\{\hollowS\geq e^{1/4}(\alpha u+\beta)\}\leq e^{-u^2/4}
\een
for all $u\geq\!\sqrt{Q_1}$. In particular, under the assumption $\inputdim\geq e\approx2.7$, it follows that the choice $u=\sqrt{4s\log\inputdim}$ satisfies $u\geq\!\!\sqrt{Q_1}$ for $s\geq1$. Straightforward calculations reveal that conditions~\eqref{eq:condA} and~\eqref{eq:condB} ensure that $e^{1/4}(\alpha u+\beta)\leq1/2$, which together with~\eqref{eq:singularVSnorm} leads to
\ba
	\Prob\lefto\{\sigma_\text{min}(\subD)\leq 1/\sqrt{2}\right\}&\leq \Prob\lefto\{\hollowS\geq1/2\right\} \nonumber\\
	&\leq \Prob\{\hollowS\geq e^{1/4}(\alpha u+\beta)\}\nonumber\\
	&\leq e^{-u^2/4} = \inputdim^{-s}.\nonumber
\ea
%
%


\section{Prior Art}
\label{app:Troppresults}

\subsection{Tropp's (M0) Model and (\Pzero)-uniqueness}
In~\cite{tropp2008} the following model was introduced.
\vspace{0.2cm}

\begin{tabular}{lcp{0.7\textwidth}}
\hline
\multicolumn{3}{c}{Model (M0) for a signal $\outputvec=\dict\inputvec$}\\
\hline%
 The dictionary	& \dict	& has coherence \coh.\\
 The vector	& \inputvec	& has nonzero entries only in the positions corresponding to the columns of a sub-dictionary \subD of \dict; furthermore, the entries of \inputvec restricted to the chosen sparsity pattern are jointly continuous random variables.\\
The sub-dictionary	&\subD	& satisfies  $\sigma_\trm{min}(\subD)\geq1/\sqrt{2}$ and has $T<\coh^{-2}/2$ columns.\\
\hline
\end{tabular}
\vspace{0.2cm}
The following theorem builds on (M0).
\begin{thm}[\!\!{\cite[Thm. 13]{tropp2008}}]\label{thm:TroppP0}
Suppose that $\outputvec=\dict\inputvec$ is a signal drawn from Model (M0). Then \inputvec is almost surely the unique vector that satisfies the constraints
\be
	\dict\inputvec=\outputvec \quad\trm{and}\quad \normzero{\inputvec}\leq T.
\een
\end{thm}

\subsection{Tropp's (M1) Model and Recovery via BP}
In~\cite{tropp2008} the following model was introduced.
\vspace{0.2cm}

\begin{tabular}{lcp{0.7\textwidth}}
\hline
\multicolumn{3}{c}{Model (M1) for a signal $\outputvec=\dict\inputvec$}\\
\hline%
 The dictionary	& \dict	& has coherence \coh.\\
 The vector	& \inputvec	& has nonzero entries only in the positions corresponding to the columns of a sub-dictionary \subD of \dict; furthermore, the phases of its nonzero entries are i.i.d. and uniformly distributed on $[0,2\pi)$ (the magnitudes need not be i.i.d.).\\
The sub-dictionary	&\subD	& satisfies $\sigma_\trm{min}(\subD)\geq1/\sqrt{2}$ and has $T<\coh^{-2}/[8(s+1)\log\inputdim]$  columns ($s\geq1$).\\
\hline
\end{tabular}
\vspace{0.2cm}

The following theorem builds on (M1).
\begin{thm}[\!\!{\cite[Thm. 14]{tropp2008}}]\label{thm:TroppBP}
Suppose that $\outputvec=\dict\inputvec$ is a signal drawn from Model (M1). Then \inputvec is the unique solution of (BP) with probability at least $1-2\inputdim^{-s}$.
\end{thm}
\vspace{0.5mm}

If the requirements of both (M0) and (M1) are satisfied, then combining Theorems 9 and 10 yields the following statement: The unique solution of \emph{both} (\Pzero) and BP applied to $\outputvec=\dict\inputvec$ is given by \inputvec with probability at least $1-2\inputdim^{-s}$. 
Note, however, that both (M0) and (M1) require the sub-dictionary \subD to have $\sigma_\trm{min}(\subD)\geq1/\sqrt{2}$.~\fref{lem:subdictionary} shows that for $\dict=[\dicta\,\,\dictb]$ and $\subD$ consisting of \na arbitrarily chosen columns of \dicta and \nb randomly chosen columns of \dictb the sub-dictionary \subD has $\sigma_\trm{min}(\subD)\geq1/\sqrt{2}$ with probability at least $1-\inputdim^{-s}$.

\bibliography{IEEEabrv,confs-jrnls,publishers,patrick}
\bibliographystyle{IEEEtran}

\end{document}